\documentclass[usenatbib,usegraphics]{mn2e}                 %
\def\kms{km\,s$^{-1}$}

\def\Nifs{$^{56}$Ni}
\def\Cofs{$^{56}$Co}
\def\Lum{erg\,s$^{-1}$}

\input{psfig.tex}

\title[Supernova 2002bo]
{Supernova~2002bo: inadequacy of the single parameter description.}

\author[Benetti et al.]
{S. Benetti$^{1}$, P. Meikle$^{2}$, M. Stehle$^{3,4}$,
G. Altavilla$^{1}$, S. Desidera$^{1}$, G. Folatelli$^{5}$,
\newauthor
A. Goobar$^{5}$, S. Mattila$^{6}$, 
J. Mendez$^{7}$, H. Navasardyan$^{1}$, A. Pastorello$^{1}$, 
\newauthor
F. Patat$^{8}$, M. Riello$^{1}$, P. Ruiz-Lapuente$^{3,7}$, 
D. Tsvetkov$^{9}$, M. Turatto$^{1}$, 
\newauthor
P. Mazzali$^{10,3}$ and W. Hillebrandt$^{3}$,
\\
$^1$INAF - Osservatorio Astronomico di Padova, vicolo
dell'Osservatorio 5, I-35122 Padova, Italy \\
$^2$Blackett Laboratory, Imperial College London, Prince Consort Road, London, SW7 2BW United Kingdom\\
$^3$Max-Planck-Institut f\"ur Astrophysik, P.O. Box 1317 D-85741 Garching, Germany\\
$^4$Universit\"ats-Sternwarte M\"unchen, Scheinerstr. 1, D-81679 M\"unchen, Germany \\
$^5$Department of Physics, Stockholm University, AlbaNova University
Center, SE--106 91 Stockholm, Sweden\\
$^6$ Stockholm Observatory, Department of Astronomy, AlbaNova
University Center, SE--106 91 Stockholm, Sweden\\
$^7$Department of Astronomy, University of Barcelona, Marti i Franques
1, E-08028 Barcelona Spain\\
$^8$European Southern Observatory, Karl-Schwarzschild-Str. 2, D-85748
Garching bei M\"unchen, Germany\\
$^{9}$Sternberg Astronomical Institute, Moscow University,
Univeritetskii pr. 13, Moscow, 119992, Russia\\
$^{10}$INAF - Osservatorio Astronomico di Trieste, Via Tiepolo, 11,
I-34131 Trieste, Italy\\
}

\date{Received ................; accepted ................}

\begin{document}

\maketitle

\begin{abstract}
We present optical/near-infrared photometry and spectra of the
type~Ia SN~2002bo spanning epochs from --13~days before maximum
$B$-band light to +102~days after.  The pre-maximum optical coverage
is particularly complete.  The extinction deduced from the observed
colour evolution and from interstellar NaID absorption is quite high
viz. E$(B-V)=0.43\pm 0.10$.  On the other hand, model matches to the
observed spectra point to a lower reddening (E$(B-V)\sim 0.30$).  In
some respects, SN~2002bo behaves as a typical "Branch normal" type~Ia
supernova (SN~Ia) at optical and IR wavelengths.  We find a $B$-band
risetime of 17.9$\pm$0.5~days, a $\Delta m_{15}$(B) of $1.13\pm0.05$,
and a de-reddened $M_B=-19.41\pm 0.42$.  However, comparison with
other type~Ia supernovae having similar $\Delta m_{15}$(B) values
indicates that in other respects SN~2002bo is unusual.  While the
optical spectra of SN~2002bo are very similar to those of SN~1984A
($\Delta m_{15}$(B) = 1.19), lower velocities and a generally more
structured appearance are found in SNe~1990N, 1994D and 1998bu.  For
supernovae having $\Delta m_{15}$(B) $>$ 1.2, we confirm the variation
of $\cal R$(SiII) 
\citep{nugent} with
$\Delta m_{15}$(B). However, for supernovae such as SN~2002bo, with
lower values of $\Delta m_{15}$(B) the relation breaks down.
Moreover, the evolution of $\cal R$(SiII) for SN~2002bo is strikingly
different from that shown by other type~Ia supernovae.  The velocities
of SN~2002bo and 1984A derived from SII 5640\AA, SiII 6355\AA\ and
CaII H\&K lines are either much higher and/or evolve differently from
those seen in other normal SNe~Ia events. Thus, while SN~2002bo and
SN~1984A appear to be highly similar, they exhibit behaviour which is
distinctly different from other SNe~Ia having similar $\Delta
m_{15}$(B) values. We suggest that the unusually low temperature, the
presence of high-velocity intermediate-mass elements and the low
abundance of carbon at early times indicates that burning to Si
penetrated to much higher layers than in more normal type Ia
supernovae. This may be indicative of a delayed-detonation explosion.

\end{abstract}
\bigskip

\section{Introduction} \label{int}

Thermonuclear (type~Ia) supernovae (SNe~Ia) are believed to originate
from the thermonuclear disruption of a white dwarf composed of carbon
and oxygen. In the favoured scenario, the white dwarf accretes mass
(mostly hydrogen) from a companion star in a binary.  However, the
identification of the progenitor type is by no means certain.
Alternative initial scenarios include the merging of two binary white
dwarfs, or the accretion of helium \citep{hille00}.  It is generally
accepted that when the degenerate mass reaches the Chandrasekhar limit
(1.4 M$_\odot$), explosive carbon ignition occurs and burning to
nuclear statistical equilibrium ensues, forming mostly radioactive
$^{56}$Ni.  Intermediate-mass nuclei, e.g. $^{28}$Si, are produced in
the outer, lower-density regions.  These elements give rise to the
typical observed spectra of SNe~Ia, which are dominated by lines of
Fe, Si and S.  Nevertheless, the details of the explosion mechanism are
still poorly understood.  For example, it is not clear whether nuclear
burning proceeds entirely in the form of a deflagration, or whether a
subsequent transition to a detonation wave occurs. Also, we do not understand
fully what determines the mass of $^{56}$Ni produced, or if events
producing the same $^{56}$Ni mass can differ in other respects.\\

It is vital that we improve our understanding of SNe~Ia both for the
insight they can provide about astrophysical processes taking place
under extreme conditions, and because of their use in the measurement
of cosmological distances.  Observational studies of SNe~Ia at high
redshifts ($z \sim 0.3-1.2$) are yielding increasingly strong evidence
that we are living in a Universe whose expansion began to accelerate
at half its present age.  This finding is commonly taken to indicate a
finite positive cosmological constant $\Lambda$ (\citet{riess98};
\citet{perl99}) i.e. a new form of energy with negative pressure 
\citep{cald98}.  However, an important
caveat is that these cosmological conclusions rely on the assumption
that the physical properties of high--z SNe~Ia are the same as those
seen locally.  But given the uncertainties in the nature of local
SNe~Ia, it is important to test the validity of this assumption.  In
order to address this fundamental question we must endeavour to
improve our physical understanding of the SN~Ia phenomenon.\\

SN~Ia theoretical models must be tested and constrained through
comparison with observed light curves and spectral evolution.  Yet
only for a few events has even moderate coverage been achieved,
especially at infrared wavelengths.  Moreover, at all wavelengths
there is a scarcity of observations during the 2--3 weeks when the SN
is still brightening. Data obtained during this time can be
particularly effective in setting tight model constraints
\citep{riess99}.  A minority of SNe~Ia are obviously peculiar
(\citet{bruno93}; \citet{macio91bg}; \citet{macio97cn}; \citet{li01}),
although their significance for the overall picture is not clear.
However, even the so-called ``normal'' SNe~Ia display differences from
one to another, e.g. in the photospheric expansion velocities deduced
from the lines of the intermediate mass elements (IME)
(\citet{branch88}; \citet{barbon90}).  Other more subtle differences
in the photospheric spectra can also be seen. \\

The desire to make decisive progress in accounting for the observed
behaviour and diversity of SNe~Ia in terms of the explosion physics
and the nature and evolution of the progenitor provided the motivation
for the recently formed European Supernova Collaboration (ESC).  This
comprises a large consortium of European groups specialising in the
observation and modelling of SNe~Ia.  The consortium is partially
funded as an EU Research Training Network.  The ESC aims to elucidate
the nature of SNe~Ia through the acquisition of high-quality
photometry and spectra for 10--12 nearby SNe~Ia.  These data will be
used to constrain state-of-the-art models for the explosion and
progenitor, also under development by the ESC.\\

Our first target, SN~2002bo in NGC~3190 (SA-LINER type), was
discovered independently by Cacella and Hirose \citep{CH} in CCD
images taken on Mar. 9.08 UT and Mar. 9.505, respectively.  It lies at
the edge of a dust lane.  Soon after discovery, SN~2002bo was
classified as a type Ia SN at an early epoch, with the discovery date
being about 2~weeks before maximum light (\shortcite{kawa},
\shortcite{stef}, \shortcite{mat} and \shortcite{chor}).  The high
expansion velocity (about 17,700 \kms) of the Si~II 6355\AA~ doublet
was particularly indicative of an early epoch.  In this paper we
describe the results of our photometric and spectroscopic monitoring
campaign for SN~2002bo, and compare the observed properties with those
of a sample of Branch-normal \citep{branchnormal} SNe~Ia. We also
modelled two of the optical spectra, the earliest one and one very
close to maximum, in order to derive some of the properties of the SN
ejecta.  We address the problem of determining the epoch of the
spectra and in particular the reddening to the SN. We have also
modelled our earliest IR spectrum both to address the amount of
primordial carbon left in the SN ejecta and to identify the
transitions present in this spectrum.

\section{Observations} \label{obs}
Spectroscopy and imaging were carried out at several sites using a
number of different telescopes and/or instruments (Tables
\ref{obs_tab} \& \ref{spec_tab}).\\

\subsection{Optical Photometry} \label{phot}
\begin{figure}
\psfig{figure=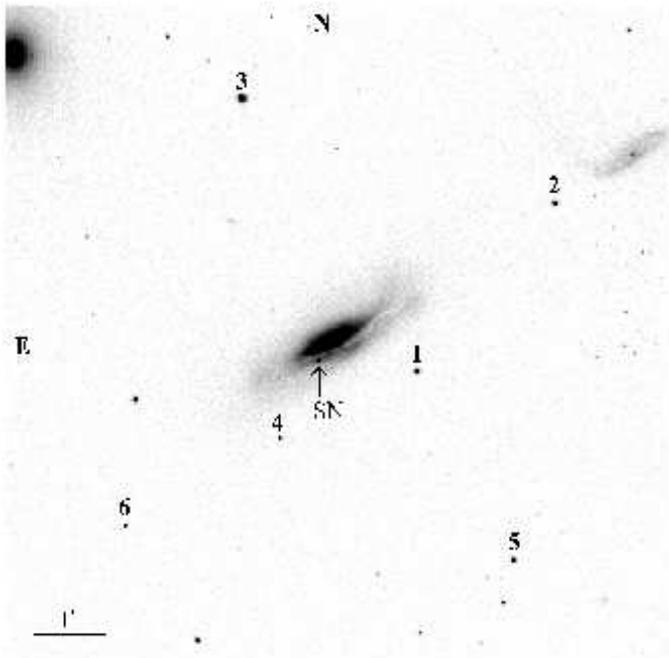,width=9cm,height=9cm}
\caption{SN~2002bo in NGC~3190 and reference stars. The image is in the
I~band and was taken with the TNG+DOLORES on 2002 May 6.} \label{sn}
\end{figure}

\begin{figure}
\psfig{figure=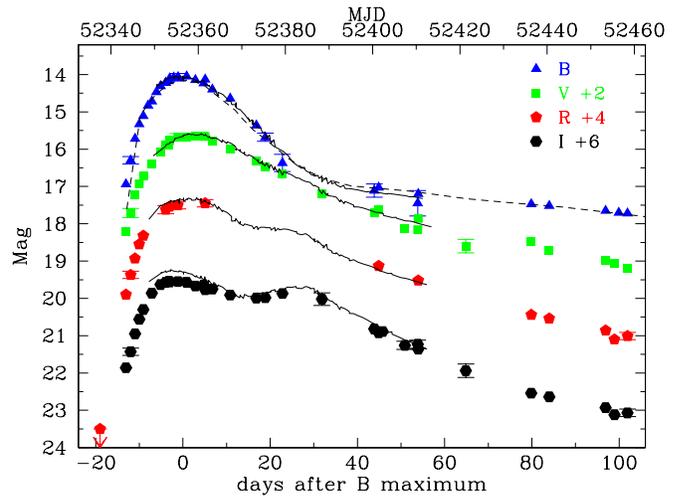,width=9.5cm,angle=-90}
\caption{B,V,R and I light curves of SN~2002bo. The R limit of
Sarneczky \& Bebesi 2002 is also shown. The solid lines represent the
B,V,R and I light curves of SN~1998bu (Suntzeff et al. 1999, Jha et
al. 1999, Hernandez et al. 2000) adjusted to the SN~2002bo distance
and reddening (see Sect. \ref{red}). A distance modulus of 30.25
(Tanvir et al.  1999) and a reddening of 0.33 (Phillips et al. 1999)
have been assumed for SN~1998bu. The dashed line is the B light curve
of SN 1984A (Barbon, Rosino \& Iijima, 1989). In this case the light
curve has been arbitrarily shifted in magnitude to fit the SN~2002bo B
maximum.}\label{phot_fig}
\end{figure}

The CCD frames were first debiased and flat-fielded in the usual
manner.  Since most of the data were obtained under non-photometric
conditions, relative photometry was derived with respect to a local
sequence of field stars (see Fig. \ref{sn}). The three photometric
nights marked in table \ref{obs_tab} (plus one VLT+FORS1 observation
of SN~2002bo on March 3rd, 2003) were used to calibrate this sequence
against Landolt standard stars \citep{land}. The magnitudes and
estimated errors of the local standards are shown in Table
\ref{seq}. These magnitudes were obtained by summing the counts
through an aperture, the size of which varied according to the seeing.
The telescope+instruments used for covering the SN~2002bo light curves
appear to define a reasonably homogeneous photometric system.  No
systematic deviations are apparent in any photometric band
(Fig. \ref{phot_fig}).  This holds even for the S70 and INT
photometric systems which have high colour terms in the colour
equations (S70 - I:$-0.40\times (R-I)$; INT - B:$+0.23\times (B-V)$,
V:$+0.135\times (B-V)$ and R:$+0.20\times (R-I)$).

\begin{table}
\caption{Magnitudes of local sequence stars identified in Fig.~1}
\label{seq}
\begin{tabular}{ccccc}
\hline
~~star~~ & B & V & R & I \\
\hline
1        &$15.19\pm0.02$ &$14.39\pm0.04$ &$13.96\pm0.05$ &$13.58\pm0.02$ \\
2        &$15.01\pm0.03$ &$14.28\pm0.06$ &$13.84\pm0.05$ &$13.52\pm0.02$ \\
3        &$13.03\pm0.09$ &$12.37\pm0.08$ &$11.98\pm0.08$ &$11.64\pm0.08$ \\
4        &$18.47\pm0.04$ &$17.18\pm0.03$ &$16.31\pm0.04$ &$15.63\pm0.04$ \\
5        &$16.31\pm0.04$ &$15.65\pm0.06$ &$15.24\pm0.07$ &               \\
6        &$18.55\pm0.06$ &$17.90\pm0.04$ &$17.51\pm0.08$ &$17.16\pm0.04$ \\
\hline
\end{tabular}
\end{table}

Ideally, one would like to remove the galaxy background by subtraction
of a galaxy ``template'' where the SN is absent. This procedure was
indeed applied to some of the S70 observations, especially those with
complex background (mostly I frames) around the SN.  However, for most
of the data a suitable template image was not available.  Therefore,
in such cases the SN magnitudes were measured using the IRAF
point-spread-function fitting task Daophot.  This procedure allows the
simultaneous fitting and subtraction of the galaxy background.  While
the pixel scales varied from one instrument to another, they were
always sufficiently small to provide good sampling of the PSFs (see
Table ~\ref{obs_tab} caption).  For cases in which the seeing is fair,
the SN is relatively bright, and its PSF well-sampled, it has been
found that this method produces results in excellent agreement with
the template subtraction method (cf. \citet{99E}).  Our results
confirm this.  The supernova magnitudes are presented in
Tab.~\ref{obs_tab}. The table lists the date (col.1), Modified Julian
Day (col.2), epoch relative to $t_{B_{max}}$ (col.3), BVRI magnitudes
with estimated internal errors in parentheses (cols.4--7), the
telescope used (col.8), and the seeing for each epoch, averaged over
the observed bands (col.9).

\begin{table*}
\caption{Photometric measurements for SN~2002bo}\label{obs_tab}
\begin{flushleft}
\begin{tabular}{lcccccclc}
\hline
 date   &  M.J.D.  &epoch$^*$  &     B       &     V         &        R      &         I      &   tel.   & avg seeing \\
        &          &(days)   &             &               &               &                &          &  (arcsec)  \\
9/3/02  & 52342.95 & --13.1 &16.94 (0.04) &  16.22 (0.03) &  15.90 (0.03) &  15.86 (0.03)  &   A1.82  & 2.4  \\
11/3/02 & 52344.01 & --12.0 &16.30 (0.10) &  15.71 (0.12) &  15.37 (0.10) &  15.43 (0.10)  &   A1.82  & 1.5  \\
12/3/02 & 52345.02 & --11.0 &15.72 (0.06) &  15.22 (0.03) &  14.93 (0.05) &  14.95 (0.06)  &   A1.82  & 2.3  \\
12/3/02 & 52345.99 & --10.0 &15.33 (0.07) &  14.93 (0.07) &  14.55 (0.05) &  14.56 (0.06)  &   A1.82  & 2.2  \\
13/3/02 & 52346.92 & --9.1  &15.11 (0.03) &  14.71 (0.03) &  14.32 (0.03) &  14.30 (0.03)  &   A1.82  & 2.6  \\
15/3/02 & 52348.02 & --8.0  &14.83 (0.03) &               &               &                &   NOT    & 1.1  \\
15/3/02 & 52348.89 & --7.1  &14.72 (0.03) &  14.39 (0.02) &               &  13.86 (0.04)  &   S70    & 2.6  \\
16/3/02 & 52349.92 & --6.1  &14.47 (0.04) &               &               &                &   NOT    & 0.9  \\
17/3/02 & 52350.89 & --5.1  &14.31 (0.06) &  14.07 (0.02) &               &  13.63 (0.05)  &   S70    & 2.0  \\
19/3/02 & 52352.04 & --4.0  &14.22 (0.03) &               &               &                &   NOT    & 3.2  \\
19/3/02 & 52352.05 & --4.0  &14.22 (0.05) &  13.88 (0.07) &  13.62 (0.11) &  13.57 (0.08)  &   A1.82  & 3.0  \\
19/3/02 & 52352.75 & --3.3  &14.17 (0.03) &  13.88 (0.02) &               &  13.54 (0.03)  &   S70    & 1.4  \\
19/3/02 & 52352.96 & --3.0  &14.10 (0.04) &  13.76 (0.03) &  13.54 (0.05) &  13.56 (0.04)  &   A1.82  & 3.7  \\
20/3/02 & 52353.89 & --2.1  &14.08 (0.02) &               &               &                &   NOT    & 3.1  \\
21/3/02 & 52354.84 & --1.2  &14.07 (0.10) &  13.68 (0.10) &  13.50 (0.10) &  13.55 (0.10)  &   A1.82  & 2.0  \\
21/3/02 & 52354.95 & --1.1  &14.08 (0.07) &               &               &                &   NOT    & 1.1  \\
23/3/02 & 52356.87 & 0.9   &14.05 (0.05) &  13.68 (0.02)  &               &  13.57 (0.02)  &   S70    & 1.8  \\
25/3/02 & 52358.85 & 2.9   &14.14 (0.06) &  13.64 (0.03)  &               &  13.67 (0.04)  &   S70    & 1.7  \\
27/3/02 & 52360.74 & 4.7   &14.23 (0.04) &  13.65 (0.02)  &               &  13.64 (0.03)  &   S70    & 1.5  \\
28/3/02 & 52361.08 & 5.1   &14.13 (0.06) &  13.66 (0.04)  &  13.45 (0.10) &  13.77 (0.04)  &   JKT    & 2.0  \\
29/3/02 & 52362.74 & 6.7   &14.40 (0.04) &  13.77 (0.02)  &               &  13.75 (0.02)  &   S70    & 1.9  \\
2/4/02  & 52366.84 & 10.8  &14.64 (0.04) &  14.00 (0.02)  &               &  13.91 (0.05)  &   S70    & 3.4  \\
8/4/02  & 52372.85 & 16.9  &15.36 (0.06) &  14.31 (0.02)  &               &  13.99 (0.05)  &   S70    & 1.5  \\
10/4/02 & 52374.89 & 18.9  &15.68 (0.11) &  14.47 (0.03)  &               &  13.98 (0.05)  &   S70    & 1.7  \\
14/4/02 & 52378.8  & 22.8  &16.37 (0.23) &  14.66 (0.03)  &               &  13.87 (0.03)  &   S70    & 1.4  \\
23/4/02 & 52387.88 & 31.9  &             &  15.20 (0.03)  &               &  14.02 (0.17)  &   S70    & 1.5  \\
5/5/02  & 52399.86 & 43.9  &17.11 (0.18) &  15.70 (0.05)  &               &  14.82 (0.06)  &   S70    & 1.6  \\
6/5/02$^{**}$$^\dag$& 52400.89 & 44.9 &17.03 (0.05) &  15.63 (0.05) &  15.13 (0.05) &  14.92 (0.05) &   TNG    & 1.9  \\
7/5/02  & 52401.82 & 45.8  &             &                &               &  14.89 (0.07)  &   S70    & 6.5  \\
12/5/02 & 52406.85 & 50.9  &             &  16.13 (0.05)  &               &  15.26 (0.11)  &   S70    & 2.6  \\
15/5/02 & 52409.83 & 53.8  &17.45 (0.34) &  16.16 (0.06)  &               &  15.23 (0.11)  &   S70    & 1.7  \\
15/5/02 & 52409.95 & 54.0  &17.20 (0.08) &  15.86 (0.06)  &  15.52 (0.03) &  15.36 (0.04)  &   A1.82  & 2.1  \\
26/5/02 & 52420.86 & 64.9  &             &  16.60 (0.18)  &               &  15.94 (0.18)  &   S70    & 4.0  \\
10/6/02 & 52435.87 & 79.9  &17.47 (0.07) &  16.47 (0.06) &  16.44 (0.08) &  16.54 (0.04)  &   A1.82  & 2.6  \\
14/6/02 & 52439.98 & 84.0  &17.52 (0.07) &  16.71 (0.06) &  16.54 (0.04) &  16.64 (0.05)  &   NTT    & 0.7  \\
27/6/02 & 52452.92 & 96.9  &17.65 (0.04) &  16.98 (0.04) &  16.86 (0.04) &  16.93 (0.09)  &   INT    & 1.1  \\
29/6/02$^\dag$ & 52454.94 & 98.9  &             &  17.06 (0.06) &  17.10 (0.04) &  17.12 (0.06)  &   JKT    & 1.4  \\
30/6/02 & 52455.92 & 99.9  &17.70 (0.05) &               &               &                &   JKT    & 2.0  \\
2/7/02$^\dag$  & 52457.94 & 101.9 &17.72 (0.08) &  17.20 (0.05) &  17.01 (0.10) &  17.07 (0.10)  &   JKT    & 2.6  \\
\hline
\end{tabular}

$*$ Epoch relative to B$_{max}$, which occurred on MJD=52356.0 = 2002 March 23.0 UT \\
$**$ We also have a $U$-band measurement at this epoch of $U=17.58\pm0.05$ \\
{\dag} Photometric night \\
A1.82 = Asiago1.82m telescope + AFOSC; pixel scale=0.473''/px\\
NOT = Nordic Optical Telescope + ALFOSC; pixel scale=0.188''/px\\
S70 = 0.70m Sternberg Astronomical Institute telescope + CCD Camera; pixel scale=1.00''/px\\
JKT = 1.0m Jacobus Kapteyn Telescope + CCD camera; pixel scale=0.330''/px\\
TNG = Telescopio Nazionale Galileo + DOLORES; pixel scale=0.275''/px\\
NTT = ESO NTT + EMMI; pixel scale=0.1665''/px\\
INT = 2.5m Isaac Newton Telescope + WFC; pixel scale=0.333''/px\\
\end{flushleft}
\end{table*}

\subsubsection{Light curves and rise times}
The B, V, R and I light curves of SN~2002bo are shown in
Fig.~\ref{phot_fig}, including the pre-discovery limit of
\shortcite{sar}. The light curves are very well sampled, spanning epochs
from just a few days after the explosion to 100 days post-maximum.  We
used a high order Legendre polynomial to fit the points around
maximum, and hence infer a B$_{max} = 14.04\pm 0.10$ mag on
$t_{B_{max}}$=MJD~$52356.0\pm 0.5$ (2002 March 23.0 UT). Thus, our
observations cover epochs between --13 and +102 days with respect to
t$_{B_{max}}$.  Likewise, we found V$_{max} = 13.58\pm 0.1$ mag, and
occurring about two days after $t_{B_{max}}$, in agreement with the
template light curves derived by \shortcite{bruno91}.  The
corresponding values for the other two bands are $R=13.49\pm 0.10$ and
$I=13.52\pm 0.10$ and occurred respectively about 1 day after and
about 2.4 days before $t_{B_{max}}$. The tendency for the $I$-band
peak to occur before t$_{B_{max}}$ has been observed in other SNe~Ia
e.g. SN~1998bu
\citep{hernandez00}. The secondary I maximum has been reached on
$=MJD~$52381.5 (e.g. $\sim 30$d after and 0.35mag below the primary
maximum). From the B light curve we derive a post-maximum decline of
$\Delta m_{15}$(B)$=1.13 \pm 0.05$.\\

Thanks to the early detection and good temporal coverage of SN~2002bo
we have been able to estimate the risetime to maximum and hence the
explosion epoch $t_0$, following \shortcite{riess99}. The main
hypothesis of this method is that the early luminosity (for epochs
$t_0 \la 10$ days) is proportional to the square of the time since
explosion.  We find $t_0(B)=$MJD $52338.7\pm 0.6$, $t_0(V)=$MJD
$52338.1\pm 0.8$, $t_0(R)=$MJD $52339.2\pm 0.8$ and $t_0(I)=$MJD
$52338.3\pm 1.0$. Thus there is reasonable agreement between the bands
as to the explosion epoch, viz.  MJD$=52338.1\pm0.5$. The B-band
risetime of 17.9$\pm0.5$~days is consistent with the Riess et
al. value of $18.7\pm 0.6$ days for a SN~Ia with the $\Delta
m_{15}(B)$ of SN~2002bo. However, our explosion epoch is in better
agreement with the $-17.6\pm0.05$ days for SN~1994D derived by
\shortcite{brunovacca} through empirical light curve fitting of B, V
and R photometry.
Finally, we note that the explosion epoch derived from modelling of
the SN~2002bo spectra (see Sect. \ref{syn}) is $-18\pm 1$ days,
which is consistent with our photometry-based value.

\subsubsection{Colour evolution}

\begin{figure}
\psfig{figure=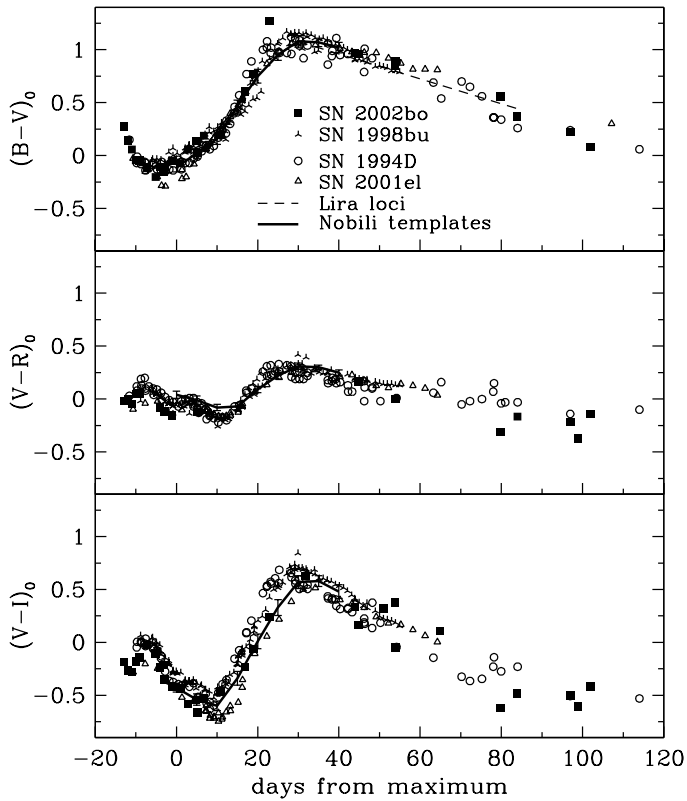,width=9.5cm,angle=0}
\caption{De-reddened $B-V$, $V-R$ and $V-I$ colour evolution of
SN~2002bo compared with that of SN 1994D (Patat et al 1996), SN1998bu
(Suntzeff et al. 1999, Jha et al. 1999, Hernandez et al. 2000),
2001el (Krisciunas et al. 2002) and the Nobili colour-curve templates
(Nobili et al. 2003). In the upper panel, the dashed line shows
the Lira loci which describes the later $B-V$ colour evolution of
unreddened SNIa.} \label{col_fig}
\end{figure}

In Fig. \ref{col_fig} we show the intrinsic $B-V$, $V-R$ and $V-I$
colours and their evolution. The $B-V$ colour was de-reddened by 0.43
mag, with corresponding values being applied to the other two colours
(see subsection 2.2). The law of \shortcite{cardelli} was used for
de-reddening.  Also shown is the colour evolution of SNe~1994D, 1998bu
and 2001el, together with the colour templates (with their intrinsic
dispersions) of \shortcite{nob} for $0-40$ days, and the loci of Lira
(\shortcite{lira}, \shortcite{phi99}). The 3 comparison SNe have
$\Delta m_{15}(B)$ values of, respectively, 1.32, 1.01 and 1.13.
Similarly to SN~1994D, two weeks before maximum the intrinsic $B-V$
colour is quite red, viz. $(B-V)_0\sim +0.28$, but it moves rapidly
blueward reaching a minimum of $-0.12$ around $-4^d$. SNe~1994D,
1998bu and 2001el display a similar behaviour, although SN~2001el
shows a slightly deeper minimum.  The curve then moves redward such
that by day +23 it appears to reach a maximum at $(B-V)_0\sim1.27$,
about 0.2 mag redder than for the comparison SNe. The curve then
follows a typical evolution to the blue, reaching a value close to
zero by $+100^d$. The Nobili et al. template nicely describes the
$B-V$ evolution in the $0-40$~day era. \\

The $V-R$ colour curve of SN~2002bo exhibits a behaviour which is very
similar to those of the comparison SNe.  Only near maximum does the
SN~2002bo curve depart from the general trend, showing values which
are about 0.10 mag. bluer. The Nobili template fails to describe the
$V-R$ evolution of our SN~Ia sample especially around phase
$+10$days. The $V-I$ colour evolution of SN~2002bo is very similar
both to those of the comparison SNe, and to the Nobili template. \\

\subsection{Near-infrared Photometry}
Near-infrared imaging (JHK) was carried out at four epochs around
maximum light at the 1.55m Carlos S\'anchez Telescope (TCS), Tenerife
using the CAIN infrared camera. This uses a 256 x 256 HgCdTe array and
was operated at a plate scale of 1.0 arcsec/pixel.  In addition a
single epoch observation at JHKs was obtained about 2 weeks after
maximum light using at the 2.5m Nordic Optical Telescope (NOT) and the
NOTCam infrared camera. This uses a 1024x1024 HgCdTe detector with a
plate scale of 0.23 arcsec/pixel. Data reduction was carried out using
standard routines within IRAF. The individual co-add frames ($\sim
10-50$ per pointing) taken with TCS have been aligned before to be combined. This
was necessary because of the inaccuracy of the telescope
tracking. Simple aperture photometry of the TCS images was carried out
using the Starlink package Gaia (Starlink User Note 214, 2003,
P.W. Draper, N.Gray \& D.S. Berry).  An aperture radius of 2
arcseconds was chosen.  This was a compromise between including as
much target signal and excluding as much galaxy background as
possible. The sky annulus had inner and outer radii of 3.4 and 5.0
arcsec respectively. However, the relatively large plate scale of the
CAIN camera together with the presence of a strong galaxy background
meant that the magnitude uncertainties, especially in the H and K
bands, are large ($\sim$0.1 to 0.2 mags). Before photometric
measurement of the NOTCam data was carried out, a smoothed background
was subtracted from the area of the supernova. This, together with the
finer plate scale of NOTCam, yielded magnitudes of higher precision.
Therefore, in order to preserve as much information as possible about
the IR evolution of the supernova, the TCS magnitudes were calibrated
using the NOTCam images via two field stars lying, respectively, about
(81''W, 7''S) and (32''E, 61''S) of the supernova.  This also meant
that the earliest TCS epoch could be used in spite of there being no
standards available for that night.  For the NOTCam calibration, use
was made of photometric standards listed in \shortcite{hunt98} and
\shortcite{persson98}. The resulting magnitudes are given in
Table \ref{ir_ph}\\

\begin{table*}
\caption{Near-infrared photometry of SN~2002bo}\label{ir_ph}
\begin{flushleft}
\begin{tabular}{ccccccc}
\hline
date &   M.J.D.&   epoch$^*$  &    J    &       H    &        K        & tel. \\
&&                 (days)   &&&& \\   
\hline
18/3/03 & 52351.5 &  -4.5 &   13.83(0.08) &  13.96(0.13) &  13.70(0.17) & TCS \\
20/3/03 & 52353.6 &  -2.4 &   13.70(0.09) &  13.87(0.21) &  13.61(0.15) & TCS \\
22/3/03 & 52355.5 &  -0.5 &   13.88(0.08) &  13.75(0.20) &  13.70(0.21) & TCS \\
24/3/03 & 52357.5 &   1.5 &   14.01(0.09) &  13.95(0.16) &  13.56(0.11) & TCS \\
5/4/03  & 52370.4 &  14.4 &   15.50(0.07) &  14.23(0.02)  & 14.12(0.05)(Ks) & NOT \\
\hline
\end{tabular}

$*$ Epoch relative to B$_{max}$, which occurred on MJD=52356.0 = 2002 March 23.0 UT \\
TCS = Carlos S\'anchez Telescope + CAIN; pixel scale = 1.0''/px \\ 
NOT = Nordic Optical Telescope + NOTCam; pixel scale = 0.23''/px \\
\end{flushleft}
\end{table*}

Within the uncertainties, the photometry shows that the IR magnitudes
of SN~2002bo evolved in a manner typical of Type~Ia
supernovae. In the J-band, where the precision is higher, we can
deduce that maximum light in this band was about 3~days before maximum
light in B.  Similar behaviour has been seen in other SNe~Ia
(e.g. SN~1998bu: \citet{hernandez00}).  In addition the rapid
fading in the J-band by +14~days is also typical. We note that a more
extensive, high precision study of the SN~2002bo IR photometric
evolution has been carried out by N. Suntzeff and K. Krisciunas at
CTIO (private communication).  To within the uncertainties, the two
sets of data are consistent. \\

\subsection{Reddening, distance and bolometric light curve} \label{red}

\shortcite{lira} and \shortcite{phi99} found a uniform colour
evolution in the $B-V$ colour of unreddened Type Ia SNe between +30
and +90 days.  By de-reddening the observed SN~2002bo $B-V$ colour
curve until it matched the Lira relation (Fig. \ref{col_fig}), we
deduced a colour excess of E$(B-V) \sim 0.47$ mag.  Only a minor
fraction of this is due to extinction in the Galaxy
(E$(B-V)_{Gal}=0.027$, \citet{sch98}), with the remaining
$\sim$0.44~mag. being due to extinction in the host galaxy.  That the
light from SN~2002bo suffered significant extinction in the host
galaxy is confirmed by strong NaID interstellar absorption lines at
the redshift of NGC~3190.  These lines have a total equivalent width
$EW=2.27\pm 0.20$\AA.  Using the relation E$(B-V)\ga 0.16\times
EW$(NaID) of \shortcite{macio02} yields E$(B-V)_{host}\ga 0.36$, in
good agreement with the value of 0.44 derived from the late
photometry.  Given the intrinsic dispersion in both the E$(B-V) -
EW$(NaID) relation and the Lira relation, we conservatively adopt
E$(B-V)=0.43\pm 0.10$. We note, however, that spectral modelling (see Sect.
\ref{syn}) points to a lower reddening value of E$(B-V)\sim 0.3$.\\

Adopting a redshift of +1405 \kms and correcting for LG infall (+208 
\kms) onto the Virgo cluster (LEDA) with an adopted Virgo distance of
15.3~Mpc \citep{freed01} we derive a distance modulus of
$\mu_B=31.67$ (21.6~Mpc) for NGC~3190.

This value is adopted throughout the paper.  Given the apparent peak
magnitudes and the reddening, we obtain absolute peak intrinsic
(i.e. de-reddened) magnitudes of $M_B=-19.41\pm 0.42$, $M_V=-19.42\pm
0.33$, $M_R=-19.18\pm 0.25$, and $M_I=-18.93\pm 0.18$ for SN~2002bo.
As before, the law of \shortcite{cardelli} was used for de-reddening.
Given the uncertainties involved, the SN~2002bo magnitudes are
consistent with the mean values given by \shortcite{gibson00} and
\shortcite{saha} for SNe~Ia with Cepheid-determined distances. The
average $\Delta m_{15}$(B) of these calibrators is $1.15\pm 0.14$
\citep{alta03}, similar to that of SN~2002bo. \\

The host galaxy of SN~2002bo, NGC~3190, belongs to the LGC~194
(Leo III) galaxy group \citep{garcia93}, which contains 16 members.
The surface brightness fluctuation method of distance determination
\citep{tonry} has been applied to two of these viz. NGC~3193 and
NGC~3226, yielding values of $32.66\pm 0.18$ and $\mu=31.86\pm 0.24$,
respectively.  Together with the distance modulus for NGC~3190 derived
above, this suggests considerable depth ($\sim$16~Mpc) in LGC~194,
and may even cast doubts upon the group membership of NGC~3193.\\

\begin{figure}
\psfig{figure=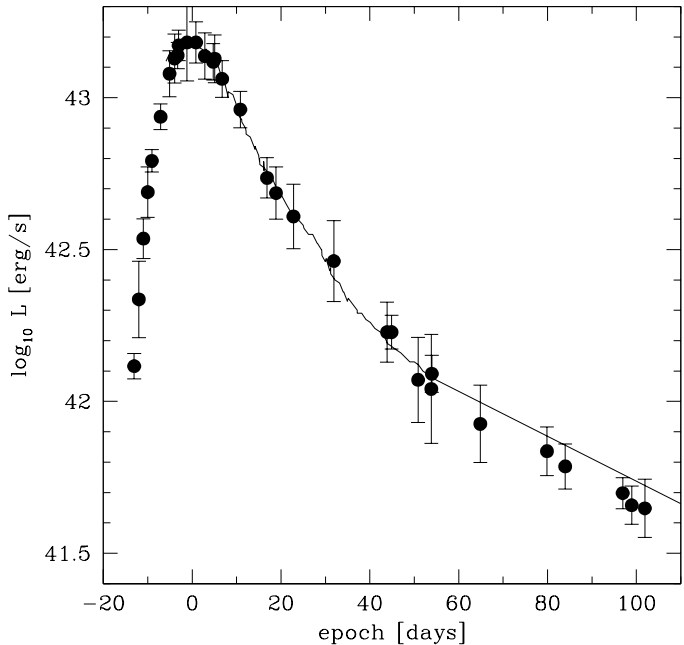,width=9.5cm,angle=0}
\caption{uvoir light curve of SN~2002bo. The solid line is the uvoir light
curve of SN 1998bu reconstructed from the published UBVRI data and
adding the UV and IR contributions given in Suntzeff (1996) for a sample
of SNIa. The error bars refer only to the photometric errors and not
to the uncertainty related to the reddening.}
\label{bolo}
\end{figure}

Integrating the fluxes in the BVRI bands, adding the IR contributions
at various epochs as discussed in Sect. \ref{IRspec}, and applying the
\shortcite{contardo} and \shortcite{nic96} corrections for the missing
U and UV contributions, respectively, we derived the uvoir luminosity
for SN~2002bo (see Fig. \ref{bolo}). For comparison the uvoir light
curve of SN 1998bu is also plotted. The similarity of the two light
curves is remarkable. SN~2002bo reached bolometric maximum on
MJD=52356 with logL=43.19~[\Lum]. The date of the maximum and the rise time
(about 17.4 days, calculated with the method described in the previous
section) closely match the maximum date and rise time found for the B
band.

A summary of the main parameter values for SN~2002bo and its host
galaxy are given in Tab. \ref{data}.

\begin{table}
\caption{Main parameter values for SN~2002bo and its host galaxy}\label{data}
\begin{tabular}{ll}
\hline
Parent galaxy           & NGC~3190      \\
Galaxy type             &  SA(s)a pec sp  LINER    $^\dag$         \\
RA (2000)               &  $10^h 18^m 06^s.51$                  \\
Dec (2000)              & $+21\degr$49'41''.7           \\
Recession velocity      [\kms] & $1405$            $^\ddag$\\
Distance modulus (H$_0=65$)& 31.67                \\
E$(B-V)$                & $0.43\pm 0.10$ \\
Offset from nucleus     &       11''.6E  ~~  14''.2S       \\
                        &                               \\
Explosion epoch (MJD)   & $52338.1\pm 0.5$ (Mar 05, 2002) \\
Date of B maximum (MJD) & $52356.0\pm0.5$ (Mar 23, 2002) \\
Magnitude at max        & $B=14.04\pm 0.10$, $V=13.58\pm 0.10$,\\
                        & $R=13.49\pm 0.10$, $I=13.52\pm 0.10$ \\
$\Delta m_{15}$(B)      & $1.13\pm0.05$      \\

\hline
\end{tabular}

{\dag} NED

{\ddag} LEDA, corrected for LG infall (208 \kms)
\end{table}

\subsection{Spectroscopy} \label{spec}
Spectroscopic observations spanned days~--12.9 to +84, with
exceptionally good temporal coverage being achieved during the
risetime, allowing us to follow day-to-day variations.  Table
\ref{spec_tab} lists the date (col.1), the Modified Julian Date
(col. 2), the epoch relative to $t_{B_{max}}$ (col.3), the wavelength
range (col.4), the instrument used (col.5), and the resolution as
measured from the FWHM of the night-sky lines (col.6).  On three
epochs ($\sim$--4, --2, --1~ days), almost contemporary spectra were
obtained at the A1.82 and NOT. These were co-added to produce the
spectra shown for these epochs.\\

\begin{figure*}
\psfig{figure=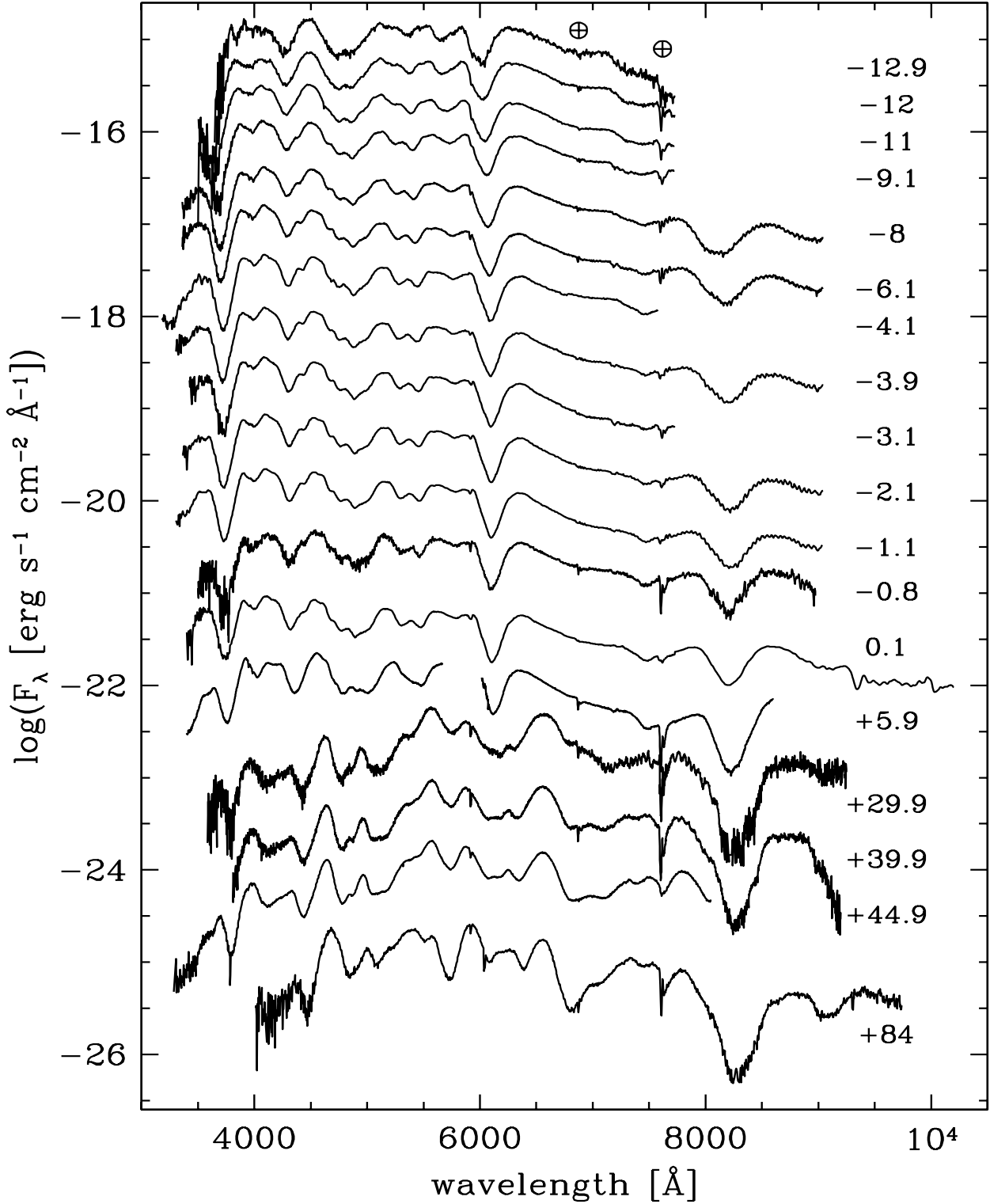,width=18cm,height=22cm,angle=0}
\caption{Spectral evolution of SN~2002bo. Wavelength is in the
observer frame. The ordinate refers to the first spectrum, and the
others have been shifted downwards by: 0.6, 1.2, 1.8, 2.4, 3, 3.6,
4.2, 4.8, 5.4, 6, 6.2, 7.4, 8, 8, 8.6, 9.2 and 9.2
respectively. Epochs are shown to the right of each spectrum. The
$\oplus$~symbols show the main telluric features}
\label{spec_evol}
\end{figure*}

The spectra were reduced following standard FIGARO or IRAF routines.
Extractions were usually weighted by the variance based on the data
values and a Poisson/CCD model using the gain and read noise
parameters. The background to either side of the SN signal was fitted
with a low order polynomial and then subtracted. Fluxing was by means
of spectrophotometric standard stars.  The flux calibration of the
optical spectra was checked against the photometry (using the IRAF
task stsdas.hst\_calib.synphot.calphot) and, where discrepancies
occurred, the spectral fluxes were scaled to match the photometry. On
nights with good observing conditions the agreement with photometry
was within 10\%.  The flux calibration of the IR spectra was checked
against $JHK$ photometry. The fluxing was adjusted where
necessary. The IR spectra on days +56/57/61 and day +85 were at later
epochs than covered by the photometry. Therefore, the IR light curves
of \shortcite{meikle00} were used to extrapolate to these ephochs. The
optical spectra are shown in Figure \ref{spec_evol}, and the IR
spectra in Figure
\ref{IRspec_evol}.\\

\subsubsection{Optical spectra}
Inspection of Figure \ref{spec_evol} shows that the early blueward
shift of the $(B-V)$ colour (Figure~3) is at least partly due to a
gradual steepening of the continuum until about $-4$~days. Similarly,
the subsequent reddening follows a gradual decrease in the continuum
slope. \\

\begin{table}
\caption{Spectroscopic observations of SN~2002bo} \label{spec_tab}
\begin{tabular}{crrrlr}
\hline
\hline
Date    & M.J.D.    & epoch$^*$ &range~~~   & tel.$^{**}$  & res.\\
        &           & (days)~   &(\AA)~~~~~ &                   &(\AA)\\
\hline
10/03/02& 52343.06  &  --12.9   & 3600-7700 & A1.82             & 25  \\
10/03/02& 52343.99  &  --12.0   & 3600-7700 & A1.82             & 25  \\
11/03/02& 52344.99  &  --11.0   & 3400-7700 & A1.82             & 25  \\
13/03/02& 52346.91  &  --9.1    & 3400-7700 & A1.82             & 25  \\
14/03/02& 52347.48  &  --8.5    & 8135-13060& UKIRT             & 25  \\
14/03/02& 52347.48  &  --8.5    &14666-25400& UKIRT             & 100 \\
15/03/02& 52348.04  &  --8.0    & 3400-9050 & NOT               & 14  \\
16/03/02& 52349.93  &  --6.1    & 3400-9050 & NOT               & 14  \\
18/03/02& 52351.85  &  --4.1    & 3200-7550 & WHT               & 2   \\
19/03/02& 52352.02  &  --4.0    & 3400-7700 & A1.82             & 25  \\
19/03/02& 52352.05  &  --3.9    & 3400-9050 & NOT               & 14  \\
19/03/02& 52352.94  &  --3.1    & 3400-7700 & A1.82             & 25  \\
20/03/02& 52353.90  &  --2.1    & 3400-9050 & NOT               & 22  \\
20/03/02& 52354.00  &  --2.0    & 3400-7700 & A1.82             & 25  \\
21/03/02& 52354.88  &  --1.1    & 3400-7700 & A1.82             & 25  \\
21/03/02& 52354.96  &  --1.0    & 3400-9050 & NOT               & 22  \\
22/03/03& 52355.18  &  --0.8    & 3200-8900 & WHT               & 12  \\
23/03/02& 52356.08  &   +0.1    & 3400-10350& A1.82             & 25  \\
28/03/03& 52361.94  &   +5.9    & 3100-8800 & WHT               & 12  \\
03/04/02& 52367.29  &  +11.3    & 8180-13390& UKIRT             & 25  \\
03/04/02& 52367.29  &  +11.3    &14725-25250& UKIRT             & 100 \\
21/04/02& 52385.40  &  +29.4    &10680-13880& UKIRT             & 25  \\
21/04/02& 52385.40  &  +29.4    &14780-25260& UKIRT             & 100 \\
21/04/02& 52385.91  &  +29.9    & 3500-9800 & INT+I             & 4   \\
23/04/02& 52387.40  &  +31.4    & 8230-10980& UKIRT             & 25  \\
01/05/02& 52395.90  &  +39.9    & 3650-9200 & INT+I             & 4   \\
06/05/02& 52400.94  &  +44.9    & 3250-8040 & TNG               & 12  \\
17/05/02& 52411.90  &  +55.9    &8155-10730 & UKIRT             & 25  \\
18/05/02& 52413.00  &  +57.0    &10730-13530& UKIRT             & 25  \\
22/05/02& 52417.00  &  +61.0    &19840-25130& UKIRT             & 100 \\
14/06/02& 52440.00  &  +84.0    & 3900-9750 & NTT               & 10  \\
15/06/02& 52440.99  &  +85.0    & 9400-16500& NTT+S             & 21  \\
\hline
\end{tabular}

* - relative to the estimated epoch of B maximum (MJD=52356.0)

** - See note to Table~2 for telescope coding plus:\\
UKIRT = United Kingdom Infrared Telescope + CGS4\\
WHT = William Herschel Telescope + ISIS\\
INT+I = Isaac Newton Telescope + IDS \\
NTT+S = ESO NTT + SOFI \\

\end{table}

\begin{figure}
\psfig{figure=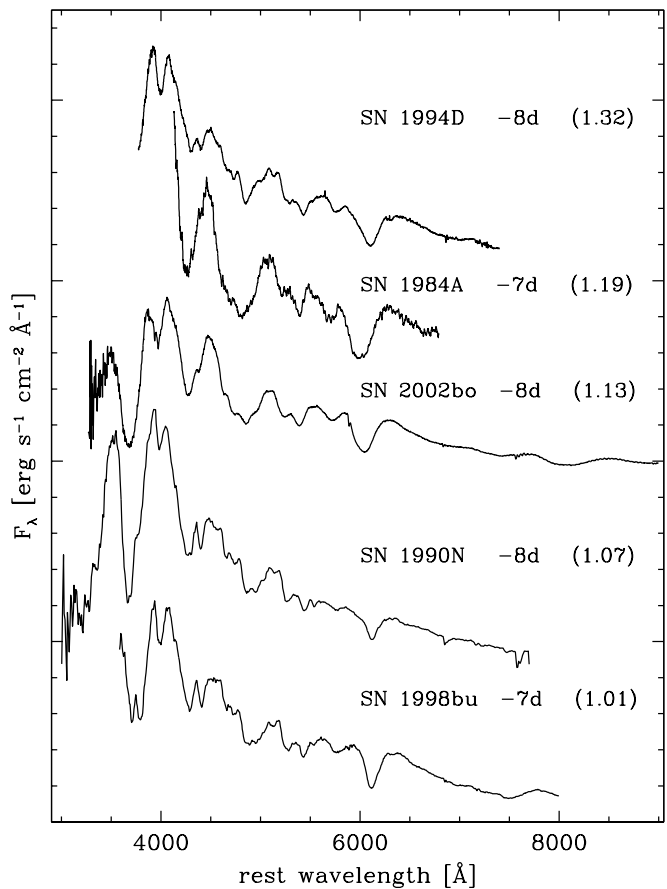,width=9.5cm,angle=0}
\caption{Comparison between spectra of SNe~Ia about a week before
maximum light arranged according to decreasing $\Delta m_{15}$(B)
(given in parenthesis). The data sources are: SN~1998bu (Hernandez et
al. 2000), SN~1994D (Patat et al. 1996), SN~1990N (Leibundgut et
al. 1991b) and SN~1984A (Benetti, 1989). The spectra have been
corrected for reddening and redshift. Epochs are shown to the right of
each spectrum.}
\label{premax}
\end{figure}

\begin{figure}
\psfig{figure=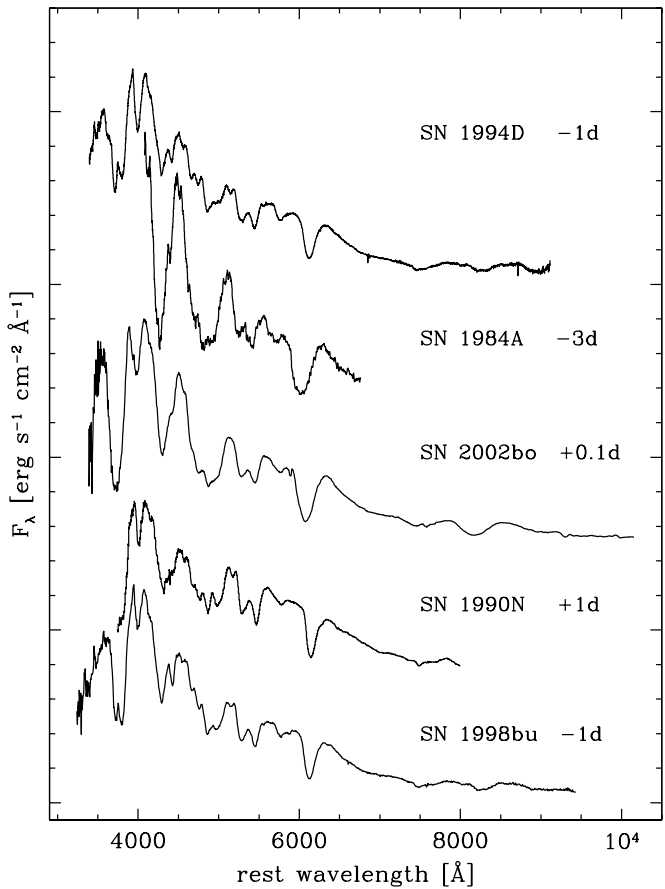,width=9.5cm,angle=0}
\caption{As for Fig. \ref{premax} but for spectra at maximum light.}
\label{max}
\end{figure}

We compare the spectra of SN~2002bo with those of SNe~1984A, 1990N,
1994D and 1998bu at about 1 week pre-maximum in Fig. \ref{premax}, and
at maximum light in Fig. \ref{max}.  All these SNe have a $\Delta
m_{15}$(B) in the range 1.01--1.32, and so in this sense can be
regarded as fairly typical.  Nevertheless, spectral differences
between the events are apparent.
At $\sim$1~week pre-maximum (Fig.~6), starting at the shortest
wavelengths we note that, unlike SNe~1990N or 2002bo, SN~1998bu shows
a double structure in the CaII H\&K absorption feature. SN~1994D shows
a similar double structure close to maximum light (Fig. \ref{max}),
and this may well have also been present in its --8d spectrum.
However, the spectral coverage stopped short of the this region.  The
prominent emission feature centered at about 4000\AA\ varies somewhat
between events owing to differing strengths in the SiII 4128,
4131\AA~ doublet. The strength of this feature appears to be proportional
to that of other SiII features in the various SNe.  In the
4000-4500\AA\ range, SN~2002bo and SN~1984A are rather similar in that
they show a strong, absorption at $\sim$4250~\AA, dominated by MgII
4481~\AA, plus a weak absorption feature at $\sim$4400~\AA\ due to
SiIII 4553, 4568\AA.  In the other three SNe, the SiIII feature is
much stronger, comparable in depth to the $\sim$4250~\AA\ absorption
which now contains both MgII 4481~\AA\ as well as a significant
contribution from FeIII (4419, 4433~\AA\ etc.), as inferred
from modelling and from the strength of the corresponding FeIII
feature near 4950\AA\ (FeIII 5074, 5127, 5156\AA).  Given the
temperature sensitivity of the SiIII line, we can immediately conclude
that both SNe 1984A and 2002bo have significantly cooler spectra than
the other three SNe shown in Fig. \ref{premax}.  In the range
4500-5000\AA, SNe 1984A and 2002bo show a strong and broad absorption
(dominated by FeII lines including multiplet 48), showing little
structure.  In contrast, the other three SNe~Ia show a much weaker
absorption, but with a greater multiplicity of small features.  Some
of these differences may be due to higher velocities in SNe~1984A and
2002bo, leading to a greater degree of line blending.  In the
5000--6000\AA\ range, while the spectra of all five SNe~Ia become more
similar to one another, we nevertheless note a somewhat larger amount of
structure in SNe 1990N, 1994D and 1998bu than in SNe~1984A and
2002bo. For example, a small dip at 5150\AA\ is present only in the
first three SNe. Longwards of 6000\AA, we find that SNe~1984A and
2002bo show SiII 6355\AA\ absorption profiles which are both broader
and more intense than in the other three SNe~Ia.\\

Turning our attention now to the maximum-light spectra SNe~Ia
(Fig. \ref{max}) we see that the differences between SNe~1984A
and 2002bo on the one hand and SNe 1990N, 1994D and 1998bu on
the other are similar to those seen at the earlier epochs.  For
example, the double structure in the CaII H\&K absorption is still
present in SN~1998bu and, as mentioned above, in SN~1994D (see
also \citet{hatano00}).  However, it remains absent in SN~2002bo.
The 4000--5000\AA\ region exhibits differences similar to those
seen at earlier times, with a strong and relatively structureless
emission feature at 4500\AA\ dominating the SNe~1984A and 2002bo
spectra, while the other three SNe~Ia show much weaker but more
structured features.  Also, the SiII 6355\AA~ absorption profiles
continue to be broader and deeper in SNe~1984A and 2002bo.

\begin{figure}
\psfig{figure=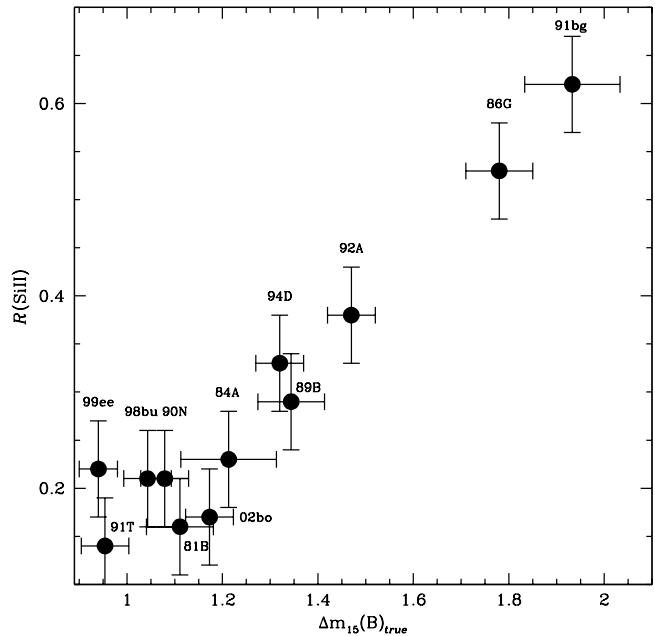,width=9.5cm,angle=0}
\caption{$\cal R$(SiII) vs. $\Delta m_{15}$(B) of the SNIa sample
of Fig. \ref{erre} plus those presented in Nugent et al. 1995. Most of
the $\Delta m_{15}$(B) have been taken from Phillips et al. (1999) but
that of SN~1999ee (Stritzinger et al 2002), SNe~2002bo and 1984A
(this paper). The $\Delta m_{15}$(B) have been corrected for the
reddening effect (Phillips et al. 1999).  The $\cal R$(SiII) of our
sample have been measured by us the others have been taken from Nugent
et al. (1995).}
\label{DR}
\end{figure}

\shortcite{nugent} found that the ratio, $\cal R$(SiII), of the
depth of the SiII 5972\AA\ and SiII 6355\AA\ absorption troughs near
maximum light is related to the speed of decline (and therefore to the
luminosity) of the event, and to the characteristic temperature of
the spectrum.  Slower, brighter decliners exhibit a smaller value of
$\cal R$(SiII) at maximum.  In Figure \ref{DR} we have plotted
maximum-light $\cal R$(SiII) vs. $\Delta m_{15}$(B) for 11 SNe.
These include our SNe~Ia sample plus those presented in
\shortcite{nugent}, and SN~1999ee (we found values of $\cal R$(SiII)
for SNe~1990N and 1994D which are slightly different from those of
\citet{nugent}: 0.21 and 0.33 vs. 0.16 and 0.29 respectively).  Figure
\ref{DR} shows that while the Nugent et al. relation holds for $\Delta
m_{15}$(B)$>$1.2, at values smaller than this it breaks down.\\

\begin{figure}
\psfig{figure=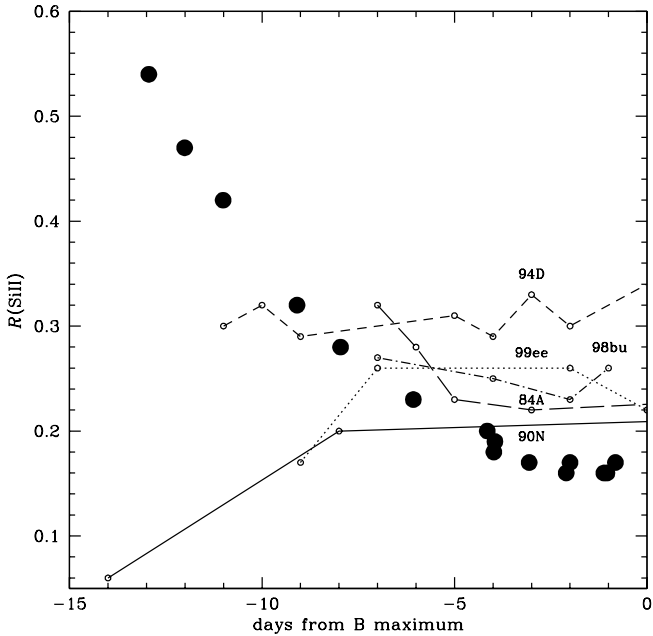,width=9.5cm,angle=0}
\caption{Evolution of $\cal R$(SiII) for a sample of SNe~Ia. Solid
circles refer to SN~2002bo data. The data sources are as in
Fig. \ref{premax}, SN~1999ee is from Hamuy et al. (2002). The $\Delta
m_{15}$(B) values for SNe~1999ee, 1998bu, 1994D, 1990N and 1984A are
0.94 (Hamuy et al.), 1.01, 1.32, 1.07 (Phillips et al 1999) and 1.19
(this paper) respectively.}
\label{erre}
\end{figure}

We also investigated the variation with epoch of the Nugent et al.
relation for SNe~1984A, 1990N, 1994D, 1998bu, 2002bo, 1999ee.
Fig. \ref{erre} shows that as we move to pre-maximum epochs the value
of $\cal R$(SiII) remains higher in the faster-declining SN~1994D
($\Delta m_{15}$(B)=1.32) than in the slower-declining SNe 1999ee,
1990N and 1998bu.  SNe~1994D and 1998bu show low amplitude variations
with epoch, without any strong trend, while SN~1990N and possibly
SN~1999ee move to lower ratios at earlier epochs.  However, SN 2002bo
exhibits a strikingly different behaviour.  $\cal R$(SiII) has a
remarkably high value of 0.54 at an epoch of $-13$d, but then
undergoes a dramatic decline, levelling out just a few days before
maximum at $\cal R$(SiII)=0.17.  This is the lowest value measured in
the entire sample at this epoch. SN~1984A also seems to follow this
trend, although it does not fall to such a low value.  The decrease of
$\cal R$(SiII) in SN~2002bo seems to track the increase of the
photospheric temperature as indicated by the decrease of the $(B-V)$
colour (see Fig. \ref{col_fig}) at these epochs.\\

\subsubsection{The expansion velocities}\label{exp_vel}

\begin{figure}
\psfig{figure=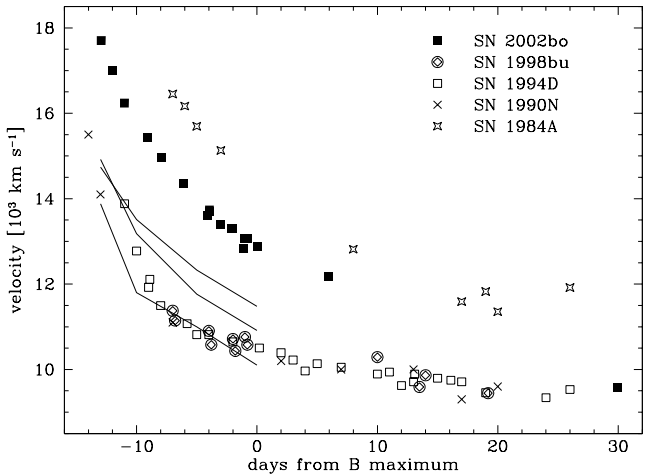,width=9.5cm,angle=-90}
\caption{Evolution of the expansion velocity deduced from the minima
of the SiII 6355\AA\ absorption for SN~2002bo, SN~1998bu (Asiago
archive, Hernandez et al. 2000), SN~1994D (Meikle et al. 1996, Patat
et al. 1996), SN~1990N (Leibundgut et al. 1991b) and SN~1984A
(Benetti, 1989). Also shown (solid lines) are the velocities predicted
by the by Lentz et al (2000) model for cases of $\times$10 (top at
epoch 0) , $\times$3 (middle) and $\times$1/3 (bottom) solar C+O layer
metallicity}
\label{SiII}
\end{figure}

Up to +5~days, SiII 6355\AA\ provides one of the deepest absorption
features.  At the earliest epochs (see Figure \ref{spec_evol}), the
profile is significantly asymmetric, owing in part to the presence of
strong NaID interstellar absorption and the smaller equivalent width
of the SiII feature at these early times.  The minimum then shifts
rapidly redwards and deepens with time as the photosphere moves into
deeper, more slowly-moving material.  This is illustrated in
Fig. \ref{SiII}, which also shows the SiII 6355\AA\ velocity evolution
for SNe~1998bu, 1994D, 1990N, 1984A.  The figure reveals clear
differences between these events.  SN~2002bo exibits an exceptionally
high velocity which decreases in a smooth, gradual manner. A similar
behaviour may be present in SN~1984A, although the data are more
sparse.  In contrast, SNe 1990N, 1994D and 1998bu show significantly
lower velocities, with a distinct break in the decline rate around
$-5$~days.  The presence of a velocity break may be related to the
fact that the velocity deduced from the SiII 6355\AA~ absorption
traces the photospheric velocity only during the era when the material
near the photosphere is Si-rich. Later, the Si absorption region
becomes more detached from the photosphere leading to a velocity which
changes more slowly with time \citep{nando94d}.\\

\noindent

\begin{figure}
\psfig{figure=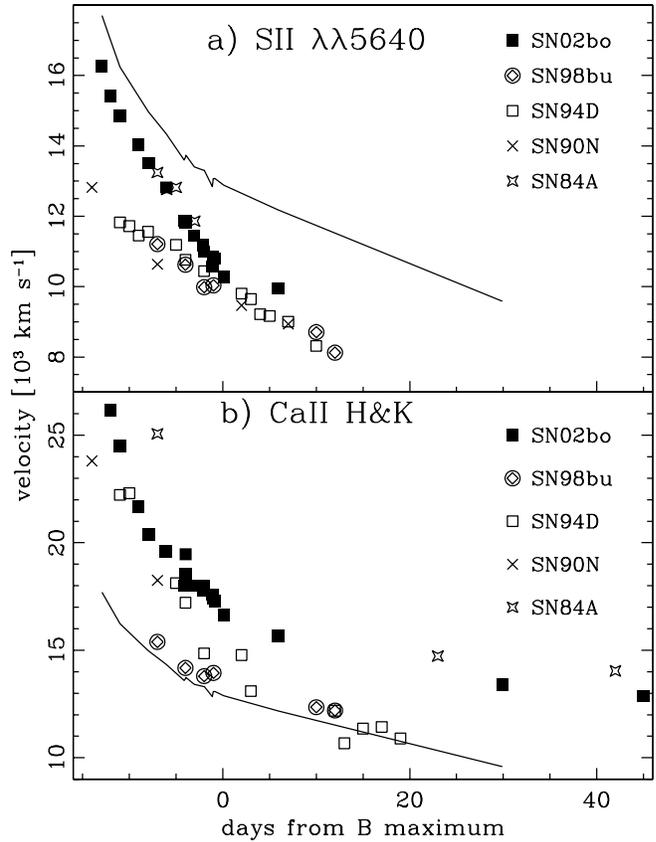,width=9.5cm,angle=0}
\caption{Evolution of the expansion velocities deduced from the minima
of the SII 5640\AA\ (a) and CaII H\&K absorptions (b). The data
sources are as in Fig. \ref{SiII} apart from CaII H \& K of SN~1984A
which are from Wegner \& McMahan (1987) and the McDonald Observatory
(unpublished). For comparison, the SN~2002bo expansion velocity from
SiII~6355\AA\ is also shown (solid line) in both panels.}
\label{Vel}
\end{figure}

\shortcite{lentz} computed emergent pre-maximum spectra for a grid of
SNIa atmospheres and argued that some of the differences between
SNe~Ia events in the blueshift of the SiII 6355\AA\ line at a given
epoch may indicate a range of metallicities in the SNIa progenitor.
In Fig. \ref{SiII} we show the Lentz model velocity predictions for
metallicity values of $\times$1/3, $\times$3 and $\times$10 solar.
While the behaviour of the lower velocity SNe~Ia is plausibly
encompassed by a metallicity range $\times$1/3 to $\times$3, a
metallicity even as high as $\times$10 solar fails to give velocities
that are anywhere near those exhibited by SN~2002bo, not to mention
the even faster SN~1984A.  This suggests that the Si observed in these
spectra is mostly the product of SN nucleosynthesis, and that the
outer layers of the ejecta of SN~2002bo do not preserve much memory of
the properties of the progenitor, as also indicated by the absence of
the CII lines discussed below. Possible causes for this are discussed
later.

We have also examined the evolution of the SII 5640\AA\ absorption
(Fig. \ref{Vel}a). This is an interesting line to study since, as it
is quite weak, we can be reasonably certain that it is always formed
close to the continuum photosphere (even allowing for possible SiIII
5740\AA\ contamination at early times \citep{paolo90n}).  Thus, it is
a valuable probe of the true photospheric velocity.  Curiously, the
pre-maximum velocity behaviour appears to divide our sample into two
groups. In SN 2002bo, the SII velocity declines from a value of over
16,000~km/s at --13~days to about 10,000~km/s at maximum.  A similar
behaviour is observed in SN~1984A, where the velocity is even slightly
higher, although here the data do not begin until about --7~days.  In
contrast, the other SNe~Ia reach only $\sim$12,500~km/s at --13 days,
but they also decline to 10,000~km/s at maximum.  This suggests rather
different conditions in the outer envelopes of the two groups.
Between 0 and +10~days, SNe~1990N, 1994D and 1998bu appear to show
roughly the same decline rate in the velocity of the SII 5640\AA\
line. Coverage of SNe~1984A and 2002bo is insufficient to reach
definite conclusions about their decline rate in this
period. Actually, at phase +8 days the SII doublet has already
disappeared in the SN~1984A spectrum.

\begin{figure*}
\psfig{figure=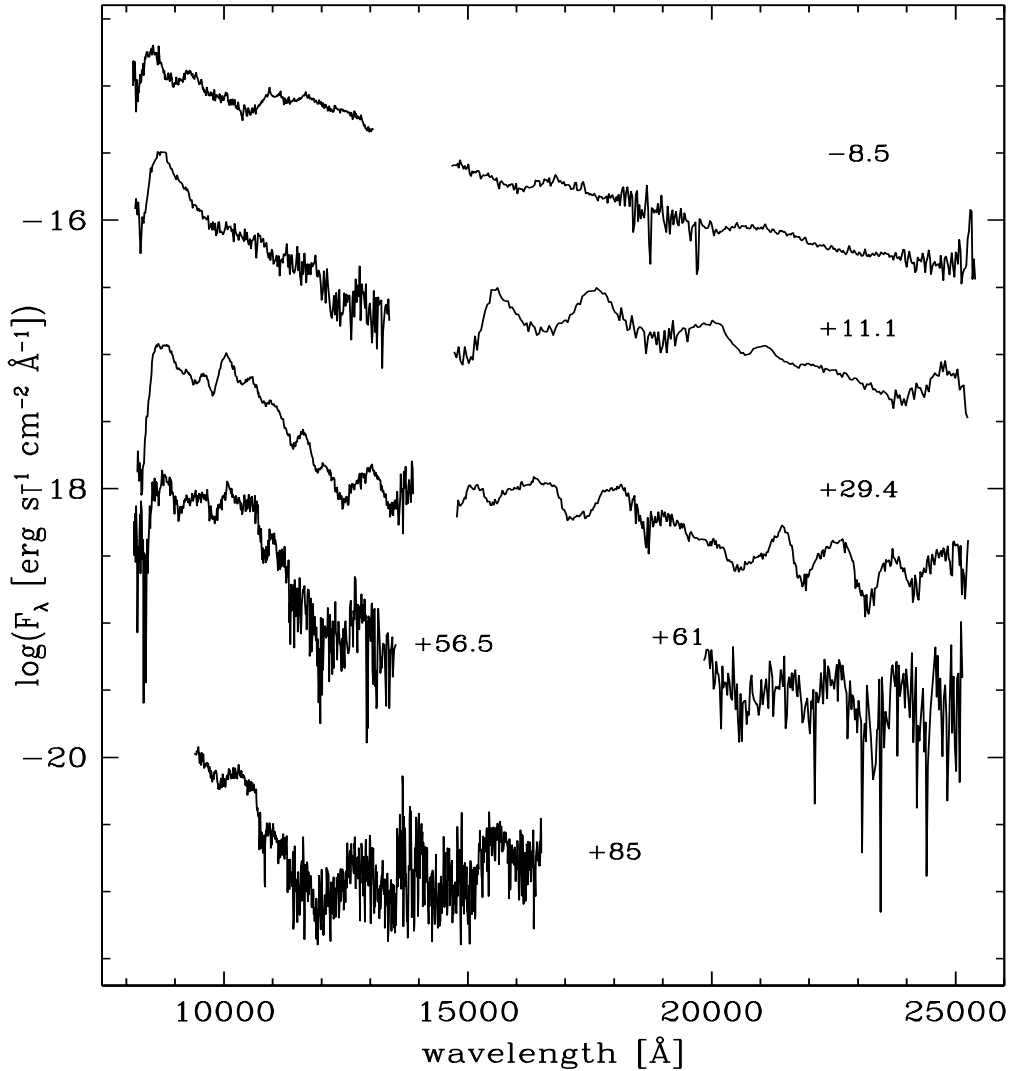,width=14cm,height=16cm,angle=0}
\caption{Spectral evolution of SN~2002bo in the infrared. Wavelength
is in the observer frame. The ordinate refers to the first spectrum,
and the others have been shifted downwards by: 1, 2.5, 2.9 and 4.4
respectively. Epochs are shown to the right of each spectrum.}
\label{IRspec_evol}
\end{figure*}

The other very prominent absorption at early times is due to the CaII
H\&K doublet.  At a given epoch this has an even higher optical depth
than SiII 6355\AA\ and so the line forms further out in the ejecta, in
higher velocity layers.  Consequently, the CaII H\&K minima exhibit
higher velocity blueshifts than those seen in contemporary SiII
6355\AA\ minima (Fig. \ref{Vel}b). Up to maximum light, the CaII H\&K
velocities of SNe~1990N, 1994D, and SN~2002bo are similar, while
SN~1998bu shows velocities which are slower by about 4000 \kms.  On
the other hand, the velocity in SN~1984A is about 5000 \kms~ higher.
By +10~days, the SN~1994D velocity has declined to values similar to
those of SN~1998bu.  On the other hand, the SN~2002bo velocity
declines more slowly, so that by +10~days its interpolated value
exceeds those of SN~1994D and SN~1998bu by $\sim$3000 \kms.  By about
day +20, the SN~1984A velocity is only 1000 \kms~ faster than the
interpolated value for SN~2002bo.\\

\subsubsection{Infrared spectra}\label{IRspec}
Inspection of Figure \ref{IRspec_evol} shows the earliest IR
spectrum (--8.5~days) to be largely featureless. A weak, complex
P-Cygni line at about 10500 \AA\ is apparent. This feature was
first noted in the early spectra of SN~1994D \citep{meikle96}, who
suggested either HeI 10830\AA\ or MgII 10926\AA\ as plausible
identifications. More detailed modelling by \shortcite{paololucy}
yielded a similarly ambiguous identification. However,
\shortcite{craig98} found that their models indicated that the
feature should be due almost entirely to MgII. Our synthetic
spectra (described below) tend to support the MgII identification.
A broad, P-Cygni profile (peak emission at $\sim$16700\AA, rest
frame) is also present, and \shortcite{craig98},
\shortcite{marion03} attribute this to SiII 16910\AA~ and MgII
16760/800\AA with the SiII dominant. Again, our synthetic spectra
confirm this. At longer wavelengths the early IR spectrum is
almost featureless except for a shallow, broad P-Cygni feature
with a peak at $\sim$20800\AA\ and attributed to SiII.

By +11~days, the spectrum has changed significantly, with a number of
prominent emission features now being present. The 10500 \AA\ feature
has vanished (similar behaviour was seen in SN~1994D,
\citep{meikle96}), while two strong, wide (FWHM $\sim 11000$\kms)
emission features have appeared at 15490 \AA~ and 17525\AA.  These are
attributed to blends of CoII, FeII and NiII (\citet{craig98},
\citet{marion03}).  The deep, characteristic J-band deficit can also
be seen \citep{spyro}.  This persists right through to the latest
spectrum at +85~d.  By one month post-maximum, three new broad
emission peaks have appeared at 21350\AA, 22490\AA\ and 23619\AA\
(rest frame), and these are attributed to Co, Ni and Si
\citep{craig98}.  They are still visible in the +56 day spectrum.\\

\begin{figure}
\psfig{figure=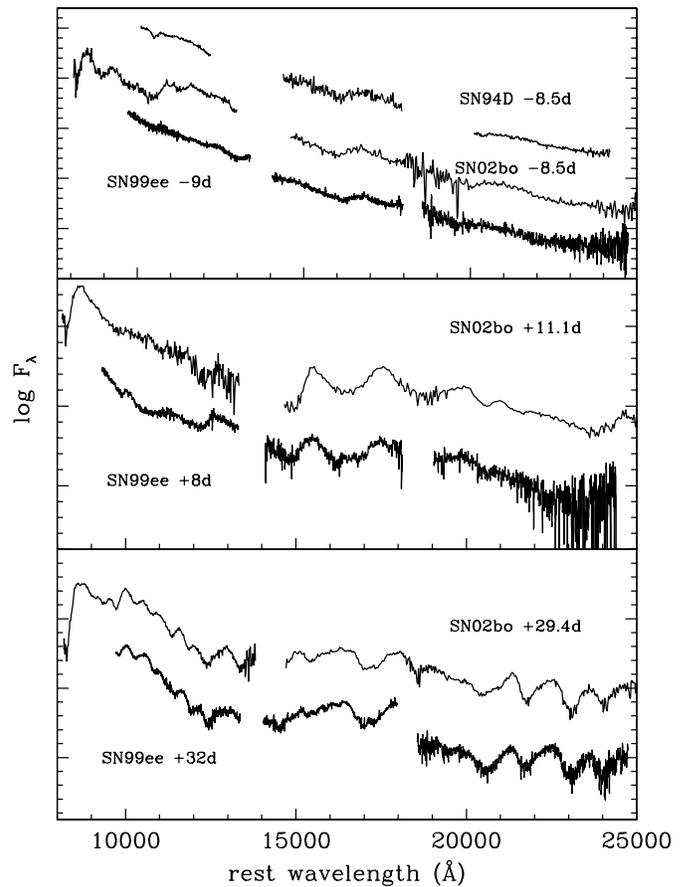,width=9.5cm,angle=0}
\caption{Comparison of SN~2002bo IR spectra with those of SN~1999ee
(from Hamuy et al. 2002) and SN~1994D for epoch --8.5d (Meikle et
al. 1996). The spectra have been corrected for the redshift of the
parent galaxies.}
\label{confIR}
\end{figure}

\shortcite{mario99ee} point out that the spectroscopic homogeneity
among Branch-normal SNe~Ia extends to the IR-domain. This is confirmed
in Figure \ref{confIR}, where we compare the IR spectra of SNe~2002bo
at $-8.5$d, +11.1d and +29.4d with those of SNe~1994D
\citep{meikle96}) and 1999ee \shortcite{mario99ee} at similar epochs.
The main difference between the three SNe~Ia is that the --8.5~d
10500\AA\ feature is absent from the SN~1999ee spectrum.  Since this
feature is also clearly visible in SN~1999by \citep{mario99ee} it
seems that SN~1999ee is peculiar in this respect.  We conclude that,
as in the optical domain, there exists some inhomogeneity among the IR
spectra of normal SNIa.\\

\begin{figure}
\psfig{figure=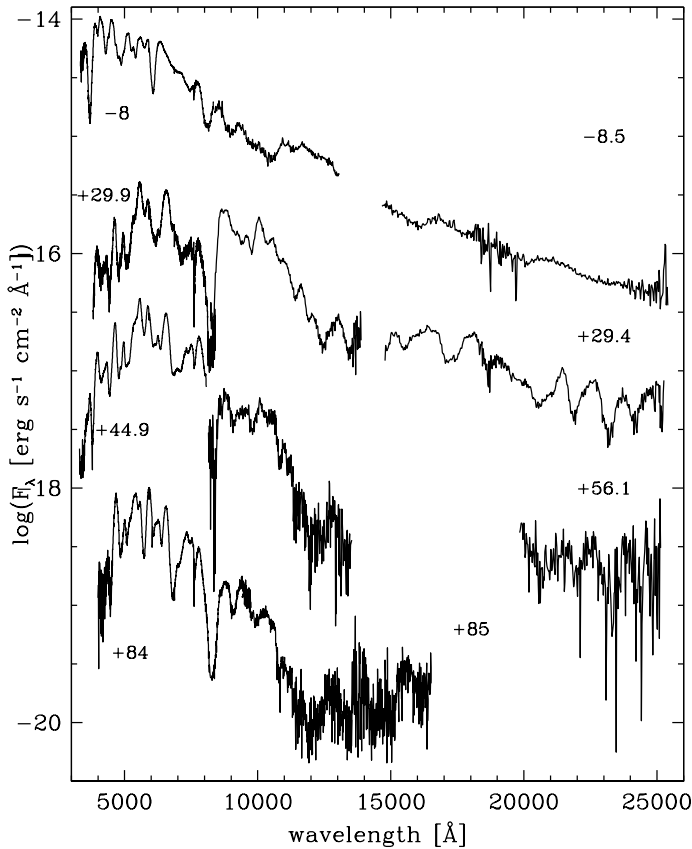,width=9.5cm,angle=0}
\caption{Opt-IR spectral evolution of SN~2002bo. Epoch from
B maximum is given for each IR and optical spectra.}
\label{IR}
\end{figure}

Figure \ref{IR} shows the overall optical+IR spectral evolution of
SN~2002bo.  This was created by combining IR and optical spectra
having similar epochs. These collages have been used to determine the
IR flux contribution for reconstructing the bolometric light curve
(see Sect. \ref{red}).  For the earliest spectrum, following
de-reddening we find that the the total flux in the IR (integrated
between 10000 and 25000~\AA) is only 5\% of the total optical flux
(integrated between 3500 and 10000 \AA). For the second one, taken
near the secondary maximum of the IR light curves, the IR contribution
rises to 18\%.  For the later two spectra the IR contribution
decreases to 6\% of the optical.  This is in close agreement with the
finding of Suntzeff (1996, 2003) who shows that more than 80\% of the
total SN~Ia uvoir flux appears in the 3000-10000\AA\ window.

\subsection{Comparison of observed and synthetic spectra}\label{syn}

In order to interpret more deeply our observations, we have computed
synthetic spectra for some of the available epochs. These models not
only provide us with information about the physical properties of the
SN ejecta, such as temperature, chemical composition, etc., but also
they can be used to verify observation-based estimates of parameters
such as reddening, distance and epoch.  We used a Monte Carlo code
originally developed by \shortcite{al85} to treat multi-line transfer
in the expanding envelopes of hot stars. This code was further
developed and adapted to SNe by \shortcite{ml93}, \shortcite{l99}, and
\shortcite{m00}. We briefly describe here the structure of the MC
code. Details can be obtained from the references given above.  The
code uses as input a model of the explosion (density v. velocity), the
emergent luminosity $L$, the epoch $t$ (time since explosion), the
estimated velocity of the photosphere $v_{ph}$ and a set of
abundances. These are treated as homogeneous above the momentary
photosphere.

The code divides the SN envelope into a number of shells, with the
thickness of each shell increasing as a function of radius.
Velocity is a continuous function of radius.  For SN ejecta we may
assume homologous expansion, $v = r/t$, where $r$ is the radius
and $t$ the time since the explosion.  Density is rescaled
according to the epoch. The temperature in the various shells is
computed assuming radiative equilibrium. At a given epoch,
temperature and density are treated as constant in each shell. The
Sobolev approximation is adopted.  Another basic assumption is
that all the radioactive decay and fast-electron energy is
deposited below a sharply-defined radius, the "photosphere"
(Schuster-Schwarzschild approximation). This energy is distributed
equally among packets, which represent ``collective photons''.
These packets are characterised by a specific frequency, and their
distribution with frequency represents the temperature at the
photosphere. The packets propagate through the envelope (i.e. the
ejecta above the photosphere) where they interact with electrons
and atoms. Interaction with electrons is treated as scattering,
while if a packet is absorbed by a line it is re-emitted in one of
the allowed downward transitions. This is selected randomly, but
weighted in proportion to the effective downward ($u \rightarrow
l$) rate of each transition. The packet is assigned the new
frequency and a random direction and the MC procedure continues
until the packet either escapes the ejecta or is absorbed back in
the photosphere. Finally, the emergent spectrum is computed using
the formal integral
\citep{l99}.\\

Here, we present and discuss synthetic spectra for two epochs.  In
view of the somewhat unusual properties of SN~2002bo (e.g. the high
velocity of several lines at early epochs), we computed models for the
earliest spectrum (epoch $\sim-13$~days) to determine whether or not
the outer abundances are peculiar.  We also computed models for a
spectrum observed near maximum light in order to check the consistency
of our results.  The density structure and initial abundances were
taken from the W7 model \citep{no84}. However, unlike the W7 model,
the composition is assumed to be uniform above the momentary
photosphere. This is done by taking the W7 abundance at the velocity
of the momentary photosphere and assuming that this is constant
throughout the outer ejecta. This procedure is repeated for each epoch
independently of the previous ones. An updated version of the code
implementing the full stratified abundance distribution is in
preparation. Keeping the density structure unchanged, the abundances
were then adjusted to improve the model match to the data. \\

\begin{figure*}
\psfig{figure=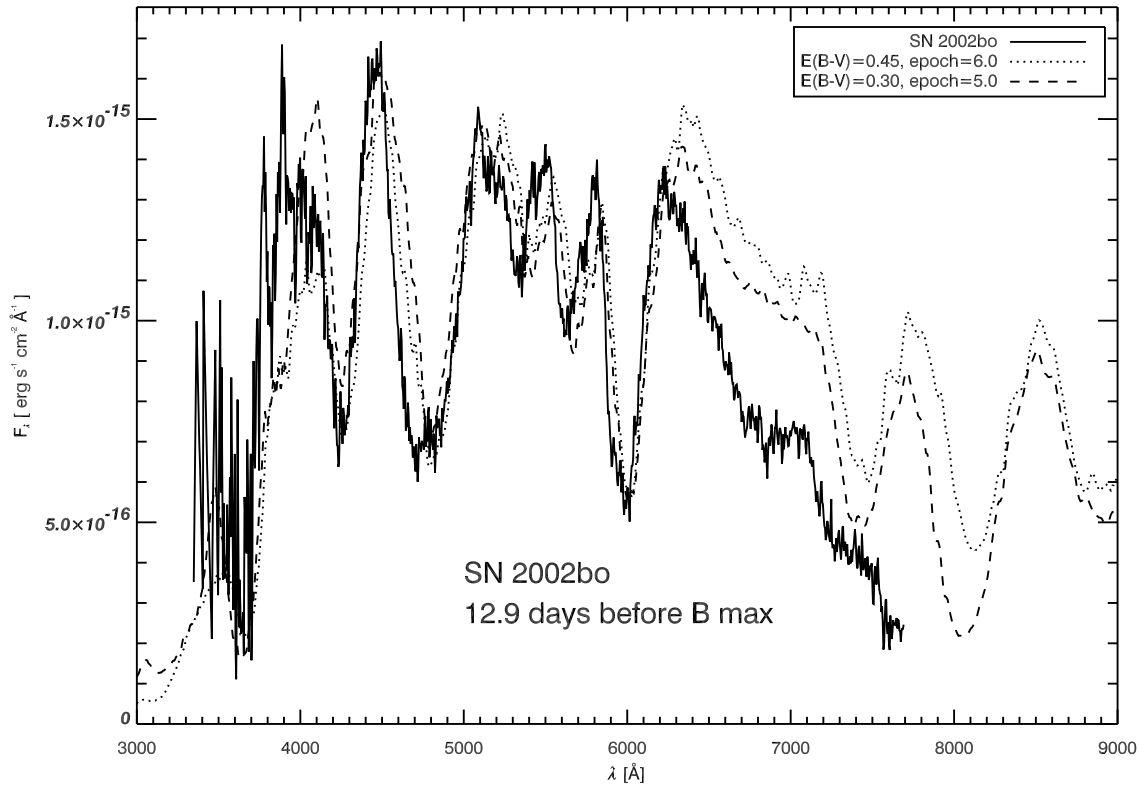,width=15cm,angle=90}
\caption{Spectrum of SN~2002bo 12.9 days before maximum. The
dotted line shows the high reddened model with an epoch of
6.0~days after explosion and the dashed line represents the model
with lower reddening at an epoch of 5.0~days after explosion.}
\label{min12_9}
\end{figure*}

We consider first the earliest spectrum (--13~days).  As indicated
above, we started with a W7-like abundance distribution. For the
envelope at this epoch, the W7 abundance is dominated by Oxygen (mass
fraction of 65\%) and a rather small contribution of Carbon (7\%).
IME are represented by Magnesium (8\%), Silicon (10\%), Sulphur (2\%)
and Calcium (2\%). Iron group elements (Titanium and Chromium 0.5\%
each, 2\% of Iron, 2\% of Nickel and 1\% of Cobalt) complete the
initial abundance set. Note that in order to reproduce observed
features an Fe abundance is adopted that is higher than the the W7
value. This is also higher than what \Cofs\ decay would allow,
indicating that a significant quantity of Fe is left over from the
progenitor. A grid of models were then computed in which radius of the
"photosphere", emergent luminosity, abundance distribution, epoch and
reddening were adjusted within a reasonable range to optimise the
match to the observed --12.9~day spectrum. Each cell of the grid
consists of a fixed epoch and reddening whilst the remaining
parameters are consequently set to optimise the fit to the data.
Although we could have included distance as one of the parameters of
the grid, we chose to concentrate on a two parameter space to simplify
our calculations. We therefore adopted a distance modulus $\mu =
31.67$, as given by the observations (Section \ref{red}). Spectra
taken at the beginning of the rise of a SN are very useful to
determine the correct epoch since luminosity changes significantly in
this period. The position of the photosphere in velocity space can be
determined fairly accurately by the position of the absorptions.
Therefore, a change in the assumed epoch translates almost directly
into a change of the photospheric radius ($R=vt$), which influences
strongly the overall temperature structure and therefore the line
depths and the ionization structure. The epoch was varied between 4
and 7 days post-explosion, and the reddening $E(B-V)$ between 0.00 and
0.45. Velocity and luminosity were adjusted to fit the overall flux
level and the position of the absorptions. The best match spectrum is
shown in Figure
\ref{min12_9}. 
\begin{figure*}
\psfig{figure=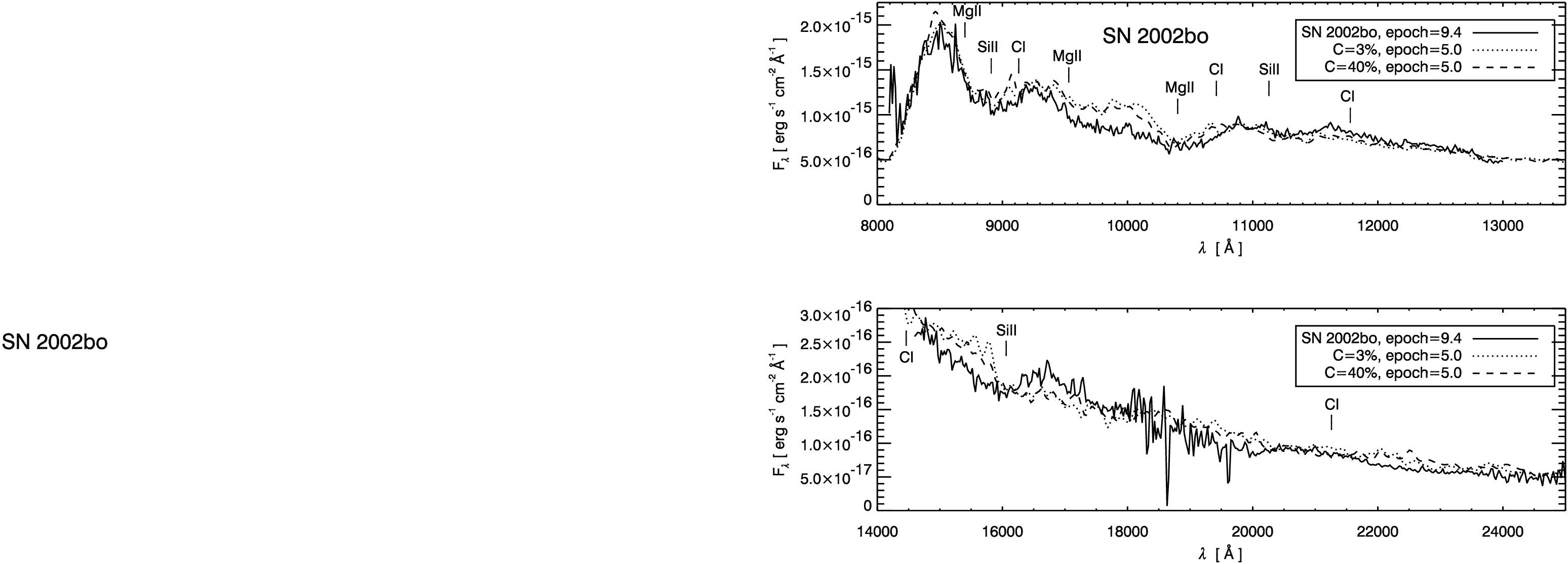,width=15cm,angle=90}
\caption{Near-infrared spectrum of SN2002bo. The --12.9~d model
has been scaled to the flux level of the observations at day --8.5
before $B_{max}$. Also shown are two versions of the low-reddening
spectral models with carbon abundances of 3\% and 40\%
respectively (see text). } \label{IR_model}
\end{figure*}
The corresponding free parameter values obtained
are: photospheric radius $=6.67\times10^{14}$~cm (corresponding to
$v_{ph} = 15,450$~\kms), log bolometric luminosity $log_{10} L =
41.94$~[\Lum], epoch = $5\pm1$ days post-explosion, $E(B-V)=0.30$
and a total mass of the envelope of $M_{tot}=0.088$~M$_\odot$. The
abundance distribution is discussed below. From this we deduce
that the maximum-light spectrum corresponds to an epoch of
$18\pm1$ days post-explosion, indicating a risetime in $B$ of the
same value. With a higher reddening of $E(B-V) = 0.45$, the best
(but poorer) match (see Fig. \ref{min12_9}) was achieved with
radius $=7.83\times10^{14}$~cm ($v_{ph} = 15,100$~\kms), $log_{10}
L = 42.13$~[\Lum] at an epoch of $6\pm1$ days post-explosion,
implying a value of $19 \pm 1$ days for the epoch of the
maximum-light spectrum and the rise time.  The abundance
distribution is discussed below. This higher-reddening match
yields a luminosity which is closer to the value of $log_{10} L
\approx 42.12$~[\Lum] inferred from Fig. \ref{bolo} for this
epoch, which could be expected, since the uvoir luminosity in fig
4 was derived using the larger reddening. If a smaller reddening
was assumed, the SN luminosity would obviously be smaller, and
comparable to that obtained from the low-reddening models here.
While the overall best match was obtained with $E(B-V) = 0.30$,
plausible matches were also obtainable with smaller reddening.
However, since the photometry and NaID absorption indicate a
higher value viz. $E(B-V) = 0.43 \pm 0.10$, we conservatively
adopt $E(B-V) = 0.30\pm 0.15$, and consider models with
both $E(B-V) = 0.45$ and 0.30.\\

The radiation temperature at the photosphere $T_R$ is found after
six iterations determining the temperature structure in the
envelope and counts for the "backwarming" effect, i.e. photons are
scattered back into the photosphere and heat it up. The values are
$T_R = 9420$~K for the low-reddening model and $T_R = 9710$~K for
the high-reddening model. Since in this temperature region there
are ionisation edges of some elements like Si and S, this
difference in temperature affects the ionization structure and
consequently the emergent spectra are significantly different. In
the low-reddening case, the Ca abundance was reduced to 1/4 of its
initial value while, for the high-reddening case, it was reduced
to 1/10. All abundances of the iron group elements are
significantly lower than in W7 ($\approx 10$\% of the initial
values), but the high-reddening model requires a higher Fe-group
abundance in order to reproduce the observed strength of the FeII lines.
This is due to the higher temperature in this model, which leads to a
lower fraction of FeII. The overall reduction of most elements caused
an increased oxygen abundance ($\ge 70$\%).

\begin{figure*}
\psfig{figure=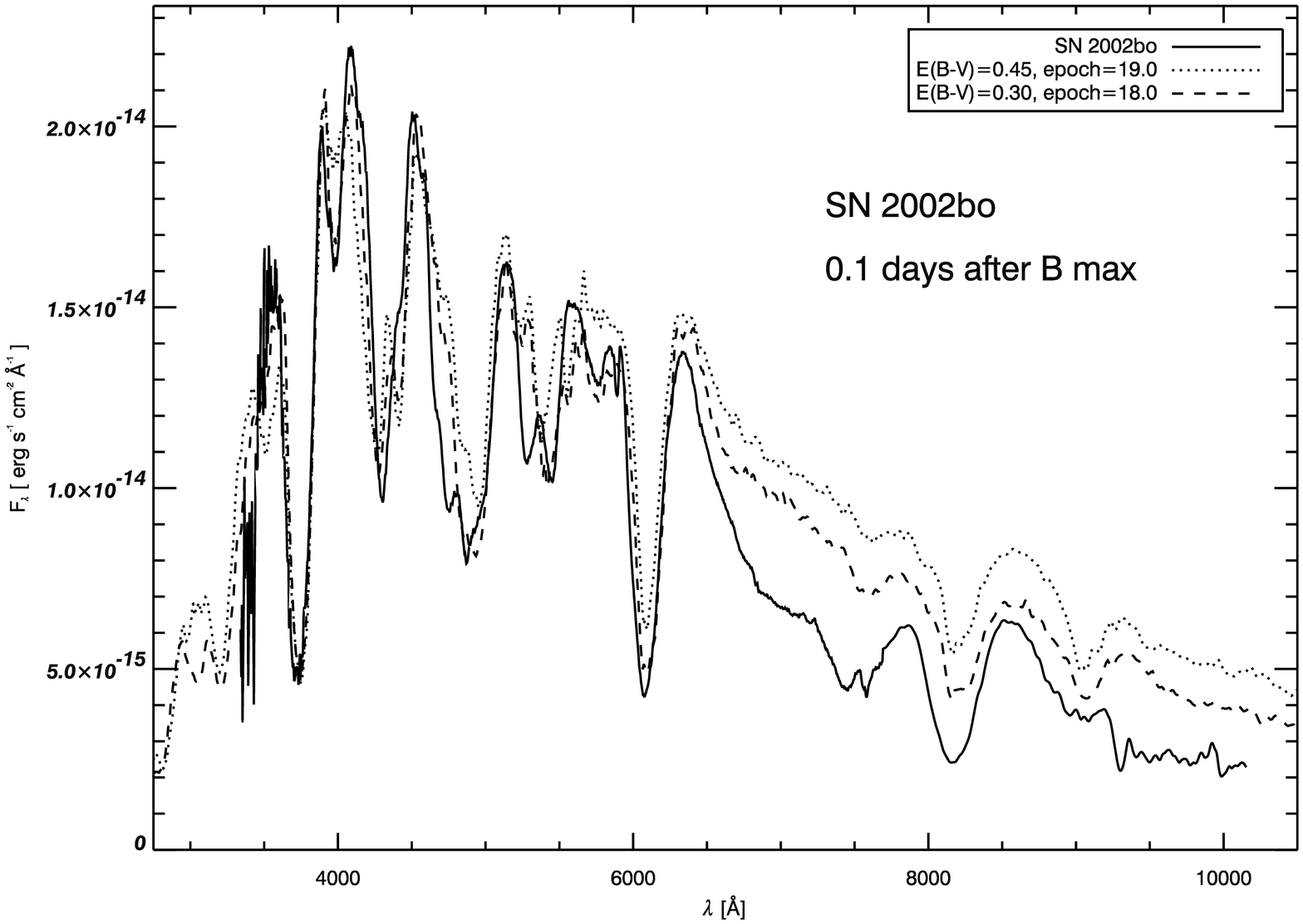,width=15cm,angle=90} \caption{Spectrum
of SN~2002bo 0.1~days after B maximum. Over-plotted are the two
models with $E(B-V) = 0.45$ (dashed) and $E(B-V) = 0.30$ (dotted
line).} \label{0_1}
\end{figure*}

We now consider in detail the model matches to the early (--13~d)
spectrum (Figure \ref{min12_9}), and examine some of the more
prominent features.  Shortward of $\sim$6500~\AA\ both models
reproduce the main features of the data quite well.  Starting at
the bluest part of the spectrum, the need for a higher temperature
in the high-reddening model is partly driven by the requirement to
reproduce the total flux in this region.  However this, in turn,
tends to increase the ratio of doubly- to singly-ionised species.
Consequently, to reproduce the deep MgII 4481~\AA\ absorbtion at
4300~\AA, a higher Mg abundance is required in the high-reddening
model.  A similar argument is relevant to the deep, broad
absorption feature at $\sim$4800~\AA, produced mostly by FeII
lines. In the high-reddening model this requires an Fe abundance
of 0.02 in the envelope, but in the low-reddening model this falls
to 0.015. We conclude that higher abundances of Mg and Fe are
required in the high-reddening model.  In contrast, to reproduce
the deep absorption feature at $\sim$5350~\AA, attributed to SII
5640~\AA, we need a somewhat higher S abundance in the {\it
low}-reddening model (0.05 compared to 0.03). We ascribe this to a
larger proportion of S recombining to the neutral state at this
lower temperature compared with the high-reddening model.  When we
consider Si, a difficulty for the high-reddening model become
apparent. In this model most of the Si is ionized to Si$^{2+}$,
making it impossible to reproduce the full depth of the 5650~\AA\
absorption, attributed to the SiII 5972~\AA\ line.  The
low-reddening model is able to reproduce the absorption features
due to SiII 5972~\AA, 6355~\AA, by invoking a relative Si
abundance of 0.12 (mass fraction) at high velocities ($v \ge
15400$~\kms). W7 predicts no Si at such high velocities from which
we may infer that extensive mixing has taken place.  However,
accepting this, we might also expect to see evidence of oxygen
mixed downward to low velocities.  That this is not observed in
the later spectra perhaps suggests that O was at least partially
burned to Si even in the outer layers.

At longer wavelengths, both models produce a large excess of flux.
This is due to the limitation of the Schuster-Schwarzschild
approximation which is used in the code. In the red/infrared part
of the spectrum, where line opacity is low, the photosphere
actually lies at a greater depth than is estimated in the code,
and consequently the model overestimates the flux.  Nevertheless,
we have carried out a comparison of model spectra with the
observations in the infrared region. Unfortunately there are no
infrared observations available at day --12.9, when the spectral
model is more applicable at such long wavelengths. The earliest
IR-spectrum is from day --8.5. Therefore, in order to examine the
8000--25,000~\AA\ region, we scaled the flux of the --12.9d
low-reddening model by $\times 1.4$ to bring it to the flux level of
the observed spectrum from day --8.5.  The IR spectrum is shown in Figure
\ref{IR_model}, together with the model spectra for two values of
the carbon abundance (see below). In general there is reasonable
agreement between the models and the observations with respect to the
overall shape of the IR spectra. In the 8000--13000~\AA\ window we
identify features due to the MgII 9217,9243~\AA\ doublet and the
10914,10915,10952~\AA\ triplet. Additionally there is a SiII feature
of minor importance at 9413~\AA. In the 15,000--18,000~\AA\ window we
attribute the broad absorption at 16,000~\AA\ in the figure to a blend
of SiII 16,906~\AA, 16,977~\AA, 17,183~\AA\ and MgII 16,760~\AA,
16,800~\AA, with the SiII feature dominating. Thus, at this early
epoch we confirm the IR identifications proposed by
\shortcite{marion03} for their spectra of other type Ia
supernovae. \shortcite{marion03} point out that the Mg~II IR features
are valuable for placing limits on the mass of unburned material. We
shall address this issue in a later paper. In the 20,000-25,000~\AA\
region line no strong features were observed in the spectrum nor
predicted by the model.  There is only a shallow, broad P-Cygni
feature at $\sim$20800\AA\ which we attribute to
SiII. \shortcite{marion03} did not cover this spectral region. A more
detailed analysis of the red/infrared part of the spectrum will be
accomplished in an upcoming paper using an improved version of the
code.\\

One reason that early-time spectra are of particular interest is
because, via the strength of CII lines in the red part of the
spectrum, they can provide interesting limits on the amount of
carbon present, and how far into the outer layers the burning
penetrated. The carbon abundance at extremely early times may even
reveal the properties of the progenitor white dwarf.  Fisher et
al. (1997) suggested that the entire `SiII' 6150~\AA\ absorption
feature in the --14~d spectrum of SN~1990N was actually CII
6578,83~\AA\ at high velocity. However, Mazzali (2001) showed that
this line could, at most, be responsible for only the red side of
the 6150~\AA\ feature, as indicated by its presence as a weak
absorption at lower velocities sitting on top of the P-Cygni
emission of the SiII line, and by the weakness of the
corresponding CII~7231,36~\AA\ line. \shortcite{branch98aq}
identified absorption features due to CII~6578,83~\AA\ and
CII~7231,36~\AA\ in the $-9$~d spectrum of SN~1998aq.  They are
also clearly visible in the $-10$~d spectrum of SN~1994D
(\citet{nando94d}, \citet{hernandez00}), although by $-8$~d they
had almost disappeared (see Fig. \ref{premax}). However, in
SN~2002bo there is no trace of these features, even in the
earliest spectra. In order to place limits on the abundance of
high velocity C implied by this negative observation, we increased
the 7\% carbon relative abundance specified by W7, to 40\%
throughout the envelope. This produced very strong
CII~6578,83~\AA\ features which were not present in the observed
spectrum.  We then decreased the carbon abundance until the
CII~6578,83~\AA\ features became fully undetectable in the
synthetic spectrum.  This occurred for a C abundance of 3\%.

As a check, we extended the modelling to 25,000~\AA\ where CI and
CII lines are found in the model's line list. These include
multiplets around 9,100~\AA, 10,680~\AA, 11,750~\AA, 14,400~\AA\
and 21,200~\AA\ plus several single lines distributed throughout
the 8000--25,000~\AA\ region. Figure \ref{IR_model} shows the IR
spectrum plus the 40\% carbon and 3\% carbon models. It can be
seen that, even with a 40\% carbon abundance, the carbon features
are barely discernable. However, the observed spectrum shows no
sign whatever of the predicted carbon features confirming that the
carbon abundance is less than 40\%. Reducing the carbon abundance
to 3\% causes the lines in the 8000--25,000~\AA\ region to become
too weak or too blended with other lines (mostly SiII or MgII) to
be detectable. We conclude that the IR spectrum is consistent with
a 3\% carbon abundance, although the actual limit here is less
stringent than in the optical region. We note that, aside from the
low abundance, the weakness of the CI lines is due to the
temperature of the SN ejecta which leads to an almost completely
ionization to CII.  The low carbon abundance is discussed in the
next section.

We turn now to the optical spectrum taken close to maximum light
(Fig. \ref{0_1}). By this epoch the photospheric radius has increased
to $1.4\times10^{15}$~cm (corresponding to $v_{ph} = 9000$~\kms) in
the low reddening model, and $=1.6\times10^{15}$~cm (corresponding to
$v_{ph} = 9740$~\kms) in the high reddening model.  The photospheric
luminosities are 43.03~[\Lum] and 43.27~[\Lum] respectively. Once
again the high-reddening model attains a higher radiation temperature
($T_R = 13420$~K) compared with the low-reddening model ($T_R =
12760$~K). Note that the temperature of the spectrum near peak
brightness is hotter than at --13 days. Such behaviour was also
observed in SN~1990N (Mazzali et al. 1993), one of the few other
SNe~Ia with very early spectral coverage. It is probably due to the
fact that the heating coming from the delayed release of radiation,
still overcomes the expansion cooling of the envelope.

Both models have a problem in reproducing the 4300--4500~\AA\
absorption, which is dominated by SiII lines. To reproduce the
depth of the absorption, we need a high Si abundance (62.1\% for
the higher reddening model and 66.7\% for the lower reddening one,
respectively) indicating that the part of the ejecta with
velocities near 10,000~\kms\ is dominated by IME. As with the
high-reddening model at the earlier epoch, the need for such a
high abundance is driven by the fact that almost all the Si is
doubly-ionised in the model. We also note that both models produce
a double structure in the trough, with the high-reddening case
being more pronounced. Curiously, while this structure is absent
from SN~2002bo, it is present in SNe 1994D and 1998bu (Fig. \ref{max}).
The persistence of a weak absorption due to SiIII 4567~\AA\ in the
model suggests that $E(B-V)$ could be even smaller than 0.30. The
high ionisation also results in the absorption of the SiII
6355~\AA\ line being slightly too shallow. However, most striking
is the fact that the SiII 5958~\AA\ absorption, while well
reproduced in the low-reddening model is completely absent in the
high reddening one.

As with the early-epoch spectrum, both models overproduce the flux
longward of $\sim$6500~\AA, with the high-reddening model being the
most discrepant.  The observed absorption at 7500~\AA\ might be
identifiable as OI~7771~\AA. (Note that the narrower 7600~\AA\ feature
is the residual of the telluric absorption.) However, even with an
unphysically high O abundance we cannot reproduce the depth of this
feature since the high temperature ionises all the neutral oxygen. The
modelled CaII absorption at $\sim$3750~\AA\ due to the 3933,68~\AA\
H\&K doublet matches the observation very well, but the absorption at
$\sim$8200~\AA\ due to the $\sim$8500~\AA\ IR~triplet is too weak even
after taking into account the offset between the spectral model and
the observed continuum. Once again the match is somewhat better in the
low-reddening model. In general, we find that the difficulties
encountered by the high-reddening model are even larger at this epoch.

We conclude that modelling of both epochs suggests a reddening value
smaller than the E$(B-V)=0.45$ derived from the Lira and NaID
relations.  This is indicated by the line ratios, line depths, overall
shape of the spectra and the model abundances.  Although distance was
not included in our grid calculation, we computed some test models
with high reddening and various distances for the --12.9~d
spectrum. Relevant parameters are very sensitive to small changes at
these early epoch \citep{1990N}. We find that reasonable results
can be obtained for a higher reddening, but this requires using a
significantly shorter distance ($\mu \la 31.00$). In order to keep a
temperature similar to the low reddening model we need to use the same
luminosity. This leads to a bolometric luminosity at maximum light of
43.00~[\Lum], similar to the luminosity of the low reddening model
(43.03~[\Lum]). So the model-derived luminosity is in any case lower
than the value 43.19~[\Lum] suggested by the observations (Section
\ref{red}). The spectral model-derived explosion epoch of $-18\pm 1$
days is consistent with the rise-time derived from our photometry
using the
\shortcite{riess99} procedure, and with the average value for SNe~Ia
given by Riess et al. It may be that some of the difficulties
encountered with the model results presented here arise from the
artificial homogenisation of the element distribution in the envelope.
The distribution of the elements throughout the ejecta will be
addressed in a separate analysis using an improved version of the MC
code, including abundance stratification.

\section{Discussion}\label{disc}

We have presented optical/near-infrared photometry and spectra of the
type~Ia SN~2002bo spanning epochs from --13~days before maximum
$B$-band light to +102~days after.  The pre-maximum optical coverage
is particularly complete. The extinction deduced from the observed
colour evolution and from interstellar NaID absorption is quite high
viz. E$(B-V)=0.43\pm 0.10$.  On the other hand, model matches to the
observed spectra point to a lower reddening (E$(B-V)\sim 0.30$). We
have been unable to resolve this reddening dichotomy.  However, the
ESC has monitored another supernova, SN~2002dj. This event exhibits
photometric and spectroscopic similarities to SN~2002bo, but suffers
from much less interstellar extinction.  Modelling of the spectra of
SN~2002dj (Pignata et al, in preparation) may help to establish the
true extinction to SN~2002bo.

In some respects, SN~2002bo behaves as a typical "Branch normal"
type~Ia supernova (SN~Ia) at optical and IR wavelengths.  We find a
$B$-band risetime of 17.9$\pm$0.5~days, a $\Delta m_{15}$(B) of
$1.13\pm0.05$, a de-reddened $M_B=-19.41\pm 0.42$, and a bolometric
maximum of $\log L=43.19$.  However, comparison with other type~Ia
supernovae having similar $\Delta m_{15}$(B) values indicates that in
other respects, SN~2002bo is unusual.  The evolution of the SN~2002bo
$(B-V)$ and $(V-R)$ colours shows some differences from that seen in
SNe~1994D, 1998bu, 2001el (see Fig. \ref{col_fig}). Moreover, while
the optical spectra of SN~2002bo are very similar to those of SN~1984A
(which has a similar $\Delta m_{15}$(B) = 1.19), lower velocities and
a generally more structured appearance are found in SNe~1990N, 1994D
and 1998bu (see also Hatano et al. 2000), whose values of $\Delta
m_{15}$(B) are only slightly smaller (SNe 1990N and 1998bu) or
slightly larger (SN~1994D).  The evolution of $\cal R$(SiII) for
SN~2002bo is strikingly different from that shown by other type~Ia
supernovae.  The SN~2002bo spectra demonstrate the existence of S at
$16,000$~km/s, Si at $>17,500$~km/s and Ca at $>26,000$~km/s. While
small amounts of primordial abundances may be present, this cannot
explain the strength of the high velocity IME spectral features.
Moreover, modelling of the SiII 5972, 6355\AA\ lines confirms the
presence of Si at velocities higher than that predicted by W7.  We
conclude that the behaviour of SN~2002bo cannot be easily related to a
single parameter description of the properties of SNe~Ia (see also
\citet{hatano00}).

The presence of high-velocity IME (Si, S, Ca) in SN~2002bo (and
SN~1984A) may be interpreted in various ways.  One possibility is
that the explosion that became SN~2002bo was more energetic than
that of the average SN~Ia, thus setting material in motion at
higher velocities.  Given that the velocities in SNe~2002bo and
1984A are about 20\% higher than in normal SNe~Ia, the kinetic
energy of the explosion would have to be about 44\% larger.
However, these high velocities are shown only by the IME.  In the
iron-group layers the Fe nebular lines have velocities comparable
to those of normal SNe~Ia, although lying to the higher velocity
side of the distribution.  Therefore, the physical difference
between SN~2002bo and more normal SNe may not be so great.  In
support of this view, we note that most of the kinetic energy is
produced by burning to Si, while the remaining burning stages to
NSE make a further, but minor contribution.

Therefore one may imagine a situation where burning to IME continues
further out into the outermost layers in SNe 2002bo and 1984A than in
other SNe~Ia. This enhanced burning may point to some form of delayed
detonation (see also \citet{lentz01}).  This scenario has some useful
consequences.  Since the IME are produced at high velocities, but at
the same time no more $^{56}$Ni is produced than in other SNe~Ia, then
this would only have a small effect on $\Delta m_{15}$, since the
shape of the light curve depends mostly on the behaviour of the line
opacity, which is dominated by Fe-group elements (Mazzali et
al. 2001).  Moreover, the amount of progenitor material (C, O)
observed in the outer layers of the SN would be greatly reduced. In
particular, if burning proceeded at a relatively low density, C, but
not O, would be burned to IME thus explaining the unusually low
abundance of carbon at the highest velocities.  It may be that the
--13~day photosphere happened to fall at the location where C but not
O had been burned to IME.  This layer has a small velocity extent, and
its location depends on the overall properties of the explosion (it is
located further out the more \Nifs\ was synthesized, see e.g. Iwamoto
et al., Fig 25).  It is therefore possible that the original, high C
abundance might still exist in layers well above the photosphere of
the --13~day spectrum, although those layers may have densities too
low for C lines to be strong once the photosphere has receded to
deeper layers. Furthermore, this picture may explain the cool
early-time temperatures indicated both by the pre-maximum Si II line
ratio and by the SiIII 4553, 4568\AA\ feature (see Sect.\ref{syn}). If
Si extends to higher levels than normal, to velocities where unburned
material is usually found, then such Si would be subject to a
much-reduced $\gamma$-ray/fast-electron flux and its temperature would
be lower than in the Si layer of more typical SNe~Ia.  There would be
a number of observable consequences of this situation: 1) at very
early times, the Si lines would be stronger than in other SNe, and
would extend to higher velocities, 2) also at very early times, the Si
line ratio would indicate a lower temperature than in other SNe,
because the Si that contributes to the lines is located further from
the $^{56}$Ni, and 3) as time goes by it might be expected that the Si
line ratio would evolve towards higher temperatures, as confirmed in
Figure 9.

The above scenario will be more severely tested by spectral models
which include abundance stratification, and by the acquisition of even
earlier spectra. Extremely early observations may even place
constraints on the progenitor composition.  In addition, the reason
for such behaviour will be explored via detailed 3D studies of the
explosion, which are currently under way.\\

Other possible explanations exist for the atypical behaviour of
SNe~2002bo and 1984A. For example, the IME produced at deeper layers
may be more efficiently mixed upwards in SNe 2002bo and 1984A than in
other SNe~Ia. This may provide an equally plausible explanation for
all the characteristics discussed above.  One way to discriminate
between the two possibilities is to look for C and O at lower
velocities - if IME have been mixed out, C and O should have been
mixed in. \\

A somewhat more exotic scenario is that SNe~2002bo and 1984A came from
more massive progenitors, such as might be produced by a binary
white-dwarf merger. In this case, however, one might expect that not
only would more IME be produced at higher velocities, but also that
more $^{56}$Ni would be synthesised.  This scenario would lead not
only to broad Fe nebular lines in the late-time spectrum, but also to
a brighter SN \citep{arnett82}, something that does not seem to be the
case here.  The nebular lines in the late-time spectrum of SN~2002bo
have widths comparable to those of other typical SNIa.

\bigskip

\noindent
{\bf ACKNOWLEDGMENTS}

We thank K. Krisciunas and N. Suntzeff for providing us with their
$JHK$ photometry of SN~2002bo prior to publication.
We also thank J.C. Wheeler for providing us an unpublished spectrum
of SN~1984A taken at McDonald Observatory.
This work is supported in part by the European Community's Human Potential
Programme under contract HPRN-CT-2002-00303, ``The Physics of Type Ia
Supernovae''.
This work is partially based on observations collected at the European
Southern Observatory, Chile (ESO N$^o$ 169.D-0670), the Italian
Telescopio Nazionale Galileo (TNG), La Palma, the Isaac Newton (INT),
Jacobus Kapteyn (JKT) and William Herschel (WHT) Telescopes of the
Isaac Newton Group, La Palma, and the United Kingdom Infrared
Telescope (UKIRT), Hawaii.  The TNG is operated on the island of La
Palma by the Centro Galileo Galilei of INAF (Istituto Nazionale di
Astrofisica) at the Spanish Observatorio del Roque de los Muchachos of
the Instituto de Astrofisica de Canarias.  The INT, JKT and WHT are
operated on the island of La Palma by the Isaac Newton Group (ING) in
the Spanish Observatorio del Roque de los Muchachos of the Instituto
de Astrofisica de Canarias.  UKIRT is operated by the Joint Astronomy
Centre on behalf of the U.K. Particle Physics and Astronomy Research
Council. Some of the data reported here were obtained as part of the
ING and UKIRT Service Programmes. Some observations were also done
within the International Time Programme ``Omega and Lambda from
supernovae and the Physics of SNe Ia explosions'' at La Palma.  For
the Nordic Optical Telescope (NOT) observations we thank J. Gorosabel,
T. Grav, T. Dahlen and G. \"Ostlin who gave up some of their
observational time. The NOT is operated on the island of La Palma
jointly by Denmark, Finland, Iceland, Norway, and Sweden, in the
Spanish Observatorio del Roque de los Muchachos of the Instituto de
Astrofisica de Canarias.
This work has made use of the NASA/IPAC Extragalactic Database (NED)
which is operated by the Jet Propulsion Laboratory, California
Institute of Technology, under contract with the National Aeronautics
and Space Administration. We have also made use of the Lyon-Meudon
Extragalactic Database (LEDA), supplied by the LEDA team at the Centre
de Recherche Astronomique de Lyon, Observatoire de Lyon.

\end{document}